\documentclass[aps,amsmath,amssymb, notitlepage, twocolumn,
nofootinbib,
superscriptaddress,
longbibliography
]{revtex4-1}
\usepackage{caption}
\usepackage{subcaption}
\usepackage{amsmath}
\usepackage{tabularx,graphicx}
\usepackage{epstopdf}
\usepackage{graphicx}
\usepackage{latexsym}
\usepackage{amssymb}
\usepackage{amsmath}
\usepackage{color, colortbl}
\usepackage{psfrag}
\usepackage{bbm}
\usepackage{bm}
\usepackage{titlesec}
\usepackage{dsfont}
\usepackage{feynmp}
\usepackage{slashed}
\usepackage{multirow}
\newcommand{\Tr}{{\rm Tr}}
\newcommand{\gcs}{{\rm{CS}}_g}

\usepackage{color}
\definecolor{darkblue}{rgb}{0.,0.,0.4}
\definecolor{darkred}{rgb}{0.5,0.,0.}
\definecolor{BlueViolet}{RGB}{138,43,226}
\definecolor{SkyBlue}{RGB}{30,144,255}
\definecolor{DarkGreen}{RGB}{0,100,0}
\usepackage[pdftex,colorlinks=true,linkcolor=darkblue,citecolor=blue,urlcolor=darkred]{hyperref}

\begin{document}

\title{Emergent QCD$_3$ Quantum Phase Transitions of Fractional Chern Insulators}

\author{Ruochen Ma}
\affiliation{%
 Perimeter Institute for Theoretical Physics, Waterloo, Ontario N2L 2Y5, Canada 
}
\affiliation{Department of Physics and Astronomy, University of Waterloo, Waterloo, Ontario N2L 3G1, Canada}
\author{Yin-Chen He}%
\affiliation{%
 Perimeter Institute for Theoretical Physics, Waterloo, Ontario N2L 2Y5, Canada 
}%

\date{\today}

\begin{abstract}
Motivated by the recent work of QED$_3$-Chern-Simons quantum critical points of fractional Chern insulators (Phys. Rev. X \textbf{8}, 031015, (2018)), we study its non-Abelian generalizations, namely QCD$_3$-Chern-Simons quantum phase transitions of fractional Chern insulators.  
These phase transitions are described by Dirac fermions interacting with non-Abelian Chern-Simons gauge fields ($U(N)$, $SU(N)$, $USp(N)$, etc.).
Utilizing the level-rank duality of Chern-Simons gauge theory and non-Abelian parton constructions, we discuss two types of QCD$_3$ quantum phase transitions.
The first type happens between two Abelian states in different Jain sequences, as opposed  to  the  QED3 transitions  between Abelian states in the same Jain sequence.
A good example is the transition between $\sigma^{xy}=1/3$ state and $\sigma^{xy}=-1$ state, which has $N_f=2$ Dirac fermions interacting with a $U(2)$ Chern-Simons gauge field.
The second type is naturally involving non-Abelian states.
For the sake of experimental feasibility, we focus on transitions of Pfaffian-like states, including the Moore-Read Pfaffian, anti-Pfaffian, particle-hole Pfaffian, etc.
These quantum phase transitions could be realized in experimental systems such as fractional Chern insulators in graphene heterostructures.

\end{abstract}

\maketitle
\tableofcontents

\section{Introduction}

Understanding phases and phase transitions is the key of condensed matter research. 
A cornerstone of the modern condensed matter research is the discovery and understanding of integer and fractional quantum Hall phases (IQHs and FQHs)~\cite{Klitzing1980,Tsui1982,Laughlin1983}, which inspires lots of interesting concepts, i.e., topological order, fractionalization, emergent gauge fields, in last few decades~\cite{wen2004quantum, fradkin2013field}.
Given the exotic properties of quantum Hall (QH) phases, it is very natural to expect that phases transitions of QH phases will also be strikingly different from conventional phase transitions.

In early days experimental relevant phase transitions of QHs would necessarily involve disorder, making it very challenging to study~\cite{jain1990scaling,kivelson1992global,wen1993transitions,Chen1993,Ye1998_transition}.
The recent experimentally advances on integer and fractional Chern insulators (ICIs and FCIs) in graphene heterostructures~\cite{ponomarenko2013cloning,hunt2013massive,spanton2018observation} brought disorder free QH phase transitions within reach~\cite{PhysRevX.8}.    
These Chern insulators (CIs) are nothing but QH phases in the presence of periodic lattice potential.
The lattice potential not only plays the role of disorder that stabilizes QH phases, but also provides extra lattice symmetries that enable new types of phase transitions~\cite{PhysRevX.8,Maissam2014_FQHtransition,YML2014_SPT,Tarun2013_QHtransition, BarkeshliYao2015}.

A recent work~\cite{PhysRevX.8} by one of us has shown that,  the phase transitions between any two Abelian QH states in the same Jain sequence~\cite{jain1989composite} are described by the QED$_3$-Chern-Simons theory, namely $N_f$ flavors of Dirac fermions interact through an emergent U$(1)$ gauge theory at Chern-Simons level $K$.
These critical theories will flow to 3D conformal field theories (CFTs), and their properties depend on the values of $N_f$ and $K$.
This family of critical theories have recently generated a surge of interest from both the condensed matter and high energy community due to the duality conjectures (see a review~\cite{senthil2019duality}).
Some of these theories may have applications to other long-standing condensed matter problem, including spin liquids in frustrated magnetism~\cite{Hastings2000,Hermele2005,Ran2007,YCH_kagomespectrum,Iqbal2016,Song2019Unifying}, high temperature superconductivity~\cite{RantnerWen2002} as well as deconfined phase transitions~\cite{deccp,deccplong,lesikav04,wang2017deconfined}.
Given its interesting properties and wide applications, there has been a large theoretical effort to study the properties of this family of critical theory~\cite{GroverFT,QED3RG,di2016quantum,di2017scaling,karthik2016scale,DyerMonopoleTaxonomy,Karthik_monopole}, but its precise properties at small $N_f$ and $K$ remain an open issue due to its strongly interacting nature.
This makes its experimental exploration exciting, in particular the \emph{entire} family of  the QED$_3$-Chern-Simons theory (with any combination of $N_f$ and $K$) is accessible experimentally at the phase transitions of QH/CI phases.

Motivated by the progress on the QED$_3$-Chern-Simons universality class of the QH/CI transitions, here we go one step further to think about phase transitions described by the QCD$_3$-Chern-Simons theory with emergent non-Abelian gauge fields.
Similar to its QED$_3$ cousin, the QCD$_3$-Chern-Simons theory also have interesting properties such as duality~\cite{hsin2016level} and generates considerable interest in past~\cite{AharonyQCD3,GiombiQCD3,AharonyUSpduality,Aharonyduality,Radicevic2016,Hui2017,zou2018field,Rhine2019}.
In this paper, we mainly focus on two types of such phase transitions that can be understood using non-Abelian parton constructions~\cite{wen1991non}. 
One type is the phase transitions involving non-Abelian FQH/FCI states~\cite{moore1991nonabelions,wen1991non,nayak,wen1999projective,read2000paired,Barkeshli_USpN,Goldman2019}, at which non-Abelian gauge field naturally emerges.
The other type is a bit more surprising as it happens between two Abelian states in different Jain sequences, as opposed to the QED$_3$-Chern-Simons critical points between Abelian states in the same Jain sequence.

The rest of the paper is organized as follows. In Section \ref{sec:review}, we first review some previous works and terminologies that are useful for our discussion.
In particular, we will discuss the condensed matter application of level-rank duality~\cite{hsin2016level}, and clarify its difference from the usual high-energy literature. 
Then in Section \ref{sec:abelian2}, we discuss the QCD$_3$-Chern-Simons transitions between Abelian states in different Jain sequences.
One example is the transition between $\sigma^{xy}=1/3$ FQH/FCI state and $\sigma^{xy}=-1$ IQH/ICI state, which has $N_f=2$ Dirac fermions coupled to a $U(2)$ Chern-Simons gauge field.
In the next two sections we turn to study non-Abelian states, in particular we focus the Pfaffian-like states (e.g. Moore-Read Pfaffian~\cite{moore1991nonabelions}, anti-Pfaffian~\cite{LeeAPf,LevinAPf}, etc.) that are close to experimental realizations~\cite{willett1987observation,dolev2008observation,banerjee2018observation,zibrov2018even,kim2019even}.
 In Section \ref{sec:parton} we provide parton constructions for these Pfaffian-like states, which are important building blocks for the following analysis on their phase transitions. 
 Some of these constructions are known, but they are scattered in different papers~\cite{wen1991non,wen1999projective,Barkeshli_USpN,zou2018field,balram2018parton}.
 In Section \ref{sec:tran} we discuss the QCD$_3$-Chern-Simons transitions of these Pfaffian-like states. Finally, we summarize our results and discuss some future directions in Section \ref{sec:sum}. The appendices contain various technical details regarding non-Abelian Chern-Simons descriptions of several Pfaffian-like states.

\section{Review of previous work\label{sec:review}}

\subsection{Chern number changing transitions of free fermion}

We start by briefly reviewing transitions between ICIs at fixed electron density and flux, since they are the building block of the later discussion.
As will be clear later, the fractionalized transitions can be formulated as the Chern number changing transitions of composite fermions or fermionic partons. 
Throughout our discussion we use the unit in which the area of unit cell is $a^2=1$ and $\frac{e^2}{\hbar}=1$, so that the Dirac flux quantum is $\phi_0=2\pi$ and the fundamental conductance is $\frac{e^2}{h}=\frac{1}{2\pi}$. However we will express the value of $\sigma^{xy}$ in units of $\frac{e^2}{h}$ so that the $2\pi$ factor can be omitted. The total Chern number of a free fermion insulator, which is equal to its Hall conductance, can be obtained by summing up the Chern numbers of occupied bands, and Chern number changing transitions arise from gap-closings in the underlying band structures. As a parameter such as the strength of the lattice potential changes, the conductance and valence bands may touch and Chern number is transferred between them. The transfer of Chern number $\Delta C$ between the two is mediated by the formation of $|\Delta C|$ Dirac cones. For a transition between two phases with Chern number $C_1$ and $C_2$ respectively, the effective theory of the critical point is
\begin{equation}
    \label{eq:freefermion}
    \mathcal L=\sum_{I=1}^{\Delta C}\bar{\psi}_I(i\partial \!\!\!/+A \!\!\!/-m)\psi_I+\frac{\bar{C}}{4\pi}A d A,
\end{equation}
where $\Delta C=C_2-C_1$ and $\bar{C}=(C_1+C_2)/2$. Here $A$ is the background electromagnetic field, measured relative to the original flux. $A d A$ is the short hand notation for the Chern-Simons term $A \wedge d A$ (or $\varepsilon_{\mu\nu\rho}A_\mu\partial_\nu A_\rho)$. 
$A$ is also referred as the probe field, as it can serve as a theoretical device to probe physical properties of the field theory such as Hall response and symmetry properties.
The mass $m$ is a phenomenological parameter whose physical meaning is the band gap. Integrating out the fermions when $m$ is finite, we obtain 
\begin{equation}
    \mathcal L=(\bar{C}+\frac{1}{2}\mathrm{sgn}(m)\Delta C)\frac{1}{4\pi}AdA,
\end{equation}
where $\mathrm{sgn}(m)=m/|m|$. We can see that this effective theory produces the desired response on either side of the transition.

It is worth noting that if $\Delta C$ is larger than $1$, it generically requires to tune more than one parameters (i.e. masses of different Dirac fermions). 
Such fine-tuning can be simply avoided in the presence of lattice symmetry~\cite{Maissam2014_FQHtransition,PhysRevX.8}. 
In particular, it was shown that the magnetic translation symmetry can protect Chern number changing by an arbitrary number~\cite{PhysRevX.8}.
Specifically, if the fermion sees a flux $p/q$ per unit cell (with $p, q$ being co-prime) the Chern number changing of fermions is enforced to be $|\Delta C|=q$.
In the rest of paper, we will always assume the Chern number changing with an arbitrary $\Delta C$ is protected by lattice symmetries. 

\subsection{Level-rank duality: gauge field versus spin gauge field}

Next we will briefly review the level-rank duality~\cite{hsin2016level} as well several terminologies that will be substantially used in the rest of the paper. 

When we are using the Chern-Simons theory to describe a topological order, it is very important to distinguish a spin gauge field from a normal gauge field: the former refers to the dynamical gauge field coupled to fermions, while the latter refers to dynamical gauge field coupled to bosons.~\footnote{If we put the system on a $\mathrm{spin}_c$ manifold, the spin gauge fields should be understood as $\mathrm{spin}_c$ connections.}
Throughout this paper we are using a superscript $^s$ to denote the spin gauge field (e.g. $U(1)^s$, $SU(2)^s$).
Aslo we denote spin gauge fields by Greek symbols ($\alpha, \beta,\cdots$), normal gauge fields are labeled by Latin letters ($a, b,\cdots$), and probe (background) fields are labeled by capital letters ($A, B, \cdots$).

Physically the difference between a spin gauge field and a gauge field comes from the subtlety that the fermions (that the spin gauge field couples to) will contribute a nontrivial sign factor to fractional statics of certain anyons.
The simplest example to appreciate this is to consider the Kalmeyer-Laughlin chiral spin liquid (i.e. $1/2$ Laughlin state of bosons) \cite{kalmeyer1987equivalence}.
The well known effective theory (also called  $K-$matrix formalism \cite{wen2004quantum}) of this state is $U(1)_2$,
\begin{equation}\label{eq:CSL_Kmatrix}
\mathcal{L} = -\frac{2}{4\pi} b d b +\frac{1}{2\pi} B db.
\end{equation}
The anynon in the above theory is semion, exchanging of which yields a phase factor $i$.
To get the response of the theory one can integrate gauge field $b$, 
\begin{equation}
\mathcal{L}_{response} = \frac{1}{8\pi} B dB + \gcs.
\end{equation}
The first term gives the $\sigma^{xy}=1/2$ electric Hall conductance, while the last term gives the $\kappa^{xy}=1$ thermal Hall conductance.
Note that $\textrm{CS}_g$ is the gravitational Chern-Simons term \cite{abanov2014electromagnetic,gromov2015framing}, which physically corresponds to the thermal Hall conductance.
The coefficient of gravitational Chern Simons term characterizes thermal Hall conductance in units of $(\pi/6)(k_B^2T/\hbar)$.

It is also known that one can use  fermionic parton approach to construct the Kalmeyer-Laughlin chiral spin liquid. 
We first start with a parton decomposition that fractionalizes a  spin$-1/2$, $\vec S = f^\dag \vec \sigma f/2$, $f^\dag=(f_\uparrow^\dag, f_\downarrow^\dag)$, and the Kalmeyer-Laughlin chiral spin liquid is realized by putting each parton $f_{\uparrow,\downarrow}$ on a $C=1$ band.
One can notice that there is a $U(1)$ gauge invariance (maximally there is a $SU(2)$ gauge invariance), so we can introduce a dynamical $U(1)$ ``gauge field" $\alpha$. After integrating out gapped fermionic partons we have,
\begin{equation}\label{eq:CSL_parton}
\mathcal{L} = \frac{2}{4\pi} \alpha d \alpha +\frac{1}{8\pi} B dB + 2\gcs.
\end{equation}
Further integrating out $\alpha$, the above theory yields the same response ($\sigma^{xy}=1/2$, $\kappa^{xy}=1$) as the Kalmeyer-Laughlin chiral spin liquid.
However, there seems to have an apparent discrepancy between Eq.~\eqref{eq:CSL_Kmatrix} and Eq.~\eqref{eq:CSL_parton} as their Chern-Simons levels are opposite. 
Naively one may conclude that Eq.~\eqref{eq:CSL_parton} has anti-semion, exchanging of which yields a phase factor $-i$.
The way to reconcile the inconsistency is to realize that $\alpha$ in Eq.~\eqref{eq:CSL_parton} is a spin gauge field that couples to fermions, while $b$ in Eq.~\eqref{eq:CSL_Kmatrix} is a gauge field couples to bosons.
The coupling with fermions will give an extra sign to the fractional statistics of anyons, which converts the anti-semion to semion.
Indeed the  equivalence between Eq.~\eqref{eq:CSL_Kmatrix} and Eq.~\eqref{eq:CSL_parton} is nothing but a known statement, namely the duality between $U(1)_2$ and $U(1)_{-2}$~\cite{hsin2016level}.
We remark that in Ref.~\cite{hsin2016level} and many other literature the duality between $U(1)_2$ and $U(1)_{-2}$ is achieved by adding a local fermion to  both sides of Lagrangian. 
In the condensed matter context it is more physical to use another version of the duality, namely we have gauge fields on the one side but spin gauge fields on the other side, abstractly written as $U(1)_2 \cong U(1)^s_{-2}$.

A formal and systematic approach to understand the relation between spin gauge fields and gauge fields is to use the level-rank (or more generally boson-fermion) duality of TQFTs~\cite{moore1989taming,hsin2016level}.
Throughout the paper, we will heavily rely  on the level-rank duality to study  parton constructions and phase transitions of FCIs/FQHs.
Before closing this section, we will discuss an example of a non-Abelian state to show the power of level-rank duality.

Again let us start with the familiar parton construction, $\vec S = f^\dag \vec \sigma f/2$, $f^\dag=(f_\uparrow^\dag, f_\downarrow^\dag)$, and we keep the maximal $SU(2)$ gauge invariance in our construction.
A topologically ordered state is realized by having $f_{\uparrow,\downarrow}$ in a $C=k$ band.
To figure out the concrete topological order, we can integrate out the fermions, yielding a Chern-Simons theory $SU(2)^s_{-k}$,
\begin{equation}\label{eq:SU2_ks}
\frac{k}{4\pi} \Tr(\alpha d\alpha + \frac{2}{3}\alpha^3) + 2k\gcs + \frac{k}{8\pi} B d B.
\end{equation}
Here $\alpha$ is a $SU(2)$ spin gauge field.
We emphasize again that the usual terminology of the Chern-Simons description of a topological order is to use the gauge field rather than the spin gauge field.
So we can further use the level-rank duality,
\begin{equation}
SU(2)_{-k}^s \cong U(k)_{2},
\end{equation}
to convert Eq.~\eqref{eq:SU2_ks} into a Chern-Simons theory in terms of a $U(k)$ gauge field, $a$,
\begin{equation}
-\frac{2}{4\pi} \Tr(a d a + \frac{2}{3}a^3)  + \frac{1}{2\pi} B d (\Tr a). 
\end{equation}
Now we are ready to read out the topological order of this  parton construction.
For example, $k=1$ reduces to the previously discussed example, the bosonic $\nu=1/2$ Laughlin described by the $U(1)_2$ Chern-Simons theory. 
When $k=2$~\cite{wen1991non}, we have the $U(2)_2$ topological order. 
We note that in some literature it was mistaken that the $k=2$ parton construction gives the $SU(2)_2$ topological order (also called bosonic Pfaffian). 

\subsection{QED$_3$-Chern-Simons universality and its parton construction}
The quantum phase transition between Jain sequence state $\sigma^{xy}=C_1/(kC_1+1)$ and $\sigma^{xy}=C_2/(kC_2+1)$ is described by the QED$_3$-Chern-Simons theory~\cite{PhysRevX.8},
\begin{align}
\mathcal L&=\sum_{I=1}^{|C_2-C_1|}\bar{\psi}_I(i\partial \!\!\!/+\alpha \!\!\!/+A \!\!\!/-m)\psi_I -\frac{k}{4\pi}b d b -\frac{1}{2\pi} b d \alpha \nonumber \\
&+\frac{C_1+C_2}{8\pi}(\alpha+A)d(\alpha+A)+\frac{C_1+C_2}{2}\textrm{CS}_g.
\label{eq:qed}
\end{align}
Here $\alpha_\mu$ and $b_\mu$ are dynamical $U(1)$ gauge fields, $A_\mu$ is the $U(1)$ background field (e.g. electromagnetic field if we consider electrons).
$k$ is an odd integer if one considers Jain sequence state of bosons, while $k$ is an even integer if one considers Jain sequence state of fermions.
The dynamical gauge field $b_\mu$ comes from the Chern-Simons flux attachment for composite fermions.
In many literatures, $b_\mu$ is integrated out, which yields a Chern-Simons term $1/4k\pi \alpha d\alpha$. 

This new family of QED$_3$-Chern-Simons quantum critical points was previously studied using the language of composite fermions, here we will reformulate it using the parton construction that is useful for the rest of this paper.
We consider a parton construction that fractionalizes a particle $c$ into $c=\varphi f$. Here $f$ is fermionic, while $\varphi$ is fermionic or bosonic depending on whether $c$ is a fermion or a boson.
Jain composite fermion state $\sigma^{xy}=C/(kC+1)$ is realized by the mean field ansatz that $\varphi$ is in a $1/k$ Laughlin state ($k$ is even or odd if $\varphi$ is bosonic or fermionic), and $f$ is in an integer quantum Hall state with Chern number $C$.
One could recognize that the parton $\varphi$ implements the Chern-Simons flux attachment, while fermionic parton $f$ can be viewed as the composite fermion.
The phase transition between two Jain sequence states can be formulated as the Chern number changing transition of $f$ parton.

\subsection{Possible experimental realization}

The theory of phase transition discussed above (and below) generally applies to many systems.
An ideal system for it as well as the QCD$_3$-Chern-Simon quantum phase transitions addressed later is the graphene heterostructures with a magnetic field and superlattice potential (e.g. a moire lattice)~\cite{ponomarenko2013cloning,hunt2013massive,spanton2018observation}.
This experimental system can be modeled as Landau levels subject to a weak superlattice potential. 
In specific, we consider the regime that the cyclotron gap between Landau levels is much larger than the superlattice potential strength. 
In this regime each Landau level splits into sub-bands, and CIs are realized by completely or partially filling some of these sub-bands.
The transitions between different FQHs and ICIs/FCIs can be achieved by tuning the lattice potential (e.g. potential strength).
Below we discuss a concrete example, namely a phase transition between the $\sigma^{xy}=\frac{1}{3}$ FQH state and $\sigma^{xy}=1$ ICI~\cite{PhysRevX.8}. 

\begin{figure}  
\begin{subfigure}[b]{0.5\textwidth}
  \includegraphics[width=\textwidth]{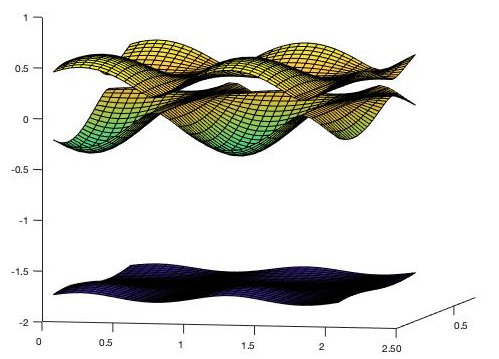} 
  \caption{}
  \end{subfigure}
  \begin{subfigure}[b]{0.5\textwidth}
  \includegraphics[width=\textwidth]{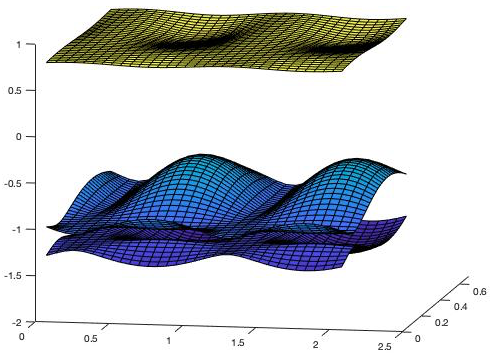} \caption{}
  \end{subfigure}
  \flushleft{ \caption{\label{fig:Chernband}
  The band structure of the LLL subject to a weak lattice potential with the form in Eq.~\eqref{eq:potential}. (a) $V_1=V_2=1$, $V_3=V_4=2.4$. The Chern number of the three bands are $C=1, 1, -1$.
  (b) $V_1=V_2=1$, $V_3=V_4=-2.4$. The Chern number of the three bands are $C=-1, 1, 1$.}}
 
\end{figure}

A general form of superlattice potential can be written as,
\begin{equation}\label{eq:potential}
\mu(\bm r) = U_0 \int d \bm r \sum_{m} ( V_m e^{i \bm r \cdot \bm G_m} n_{\bm r} +h.c.),
\end{equation}
where $G_m$ are reciprocal vectors of the superlattice.
A square lattice potential, for example, has $\bm G_1 = \frac{2\pi}{a}(1, 0)$,  $\bm G_2 = \frac{2\pi}{a}(0, 1)$, $\bm G_3 = \frac{2\pi}{a}(1, 1)$,  $\bm G_4 = \frac{2\pi}{a}(1, -1)$, and $V_1=V_2$, $V_3=V_4$, where $a$ is the lattice constant. 
We consider flux density $\phi=\frac{3}{2}$ (in units of Dirac flux quanta) and electron density $n=\frac{1}{2}$ per unit cell of the superlattice. 
Without the superlattice potential, the  lowest Landau level (LLL) is $1/3$ filled yielding the $\nu=1/3$ FQH state.
Once the superlattice potential is implemented, the LLL splits into three subbands, with a band gap $\Delta\propto U_L$. 
In the limit of strong lattice potential (i.e. potential strength much larger than the interaction strength), the lowest sub-band will be completely filled realizing a ICI state.
The Chern number will depend on the form of lattice potential.
For example, if we take $V_1=V_2=1$, $V_3=V_4=2.4$, the lowest band will have $C=1$ (see Fig.~\ref{fig:Chernband} for detailed band structure).
As one tunes the amplitude of the superlattice potential, a phase transition can be realized between $\sigma^{xy}=\frac{1}{3}$ state and $\sigma^{xy}=1$ state. 
The critical theory is described by Eq. (\ref{eq:qed}), in which $C_1=1$, $C_2=-1$ and $k=2$.

An interesting question is what will happen if we swap the order the the three subbands such that the lowest one has Chern number $C=-1$. This can be achieved by reversing the sign of the lattice potential strength used above, as shown in Fig.~\ref{fig:Chernband}. Now one obtains a transition between $\nu=\frac{1}{3}$ FQH state and $C=-1$ ICI state. 
This transition does not belong to the previously introduced QE$\mathrm{D}_3$-Chern-Simons universality. 
Instead, an non-Abelian gauge field will emerge at the quantum critical point as we will discuss in the next section.

\section{Emergent QCD$_3$ between Abelian states\label{sec:abelian2}}

\subsection{Transition between $\sigma^{xy}=1/3$ state and $\sigma^{xy}=-1$ state}

$\sigma^{xy}=1/3$ and $\sigma^{xy}=-1$ state belong to different Jain sequences ($\sigma^{xy}=C/(kC+1)$): the former has $k=2, C=1$ while the latter has $k=-2, C=1$.
Hence their transition falls beyond the QED$_3$-Chern-Simons universality. 
Below we show that their transition can be naturally captured by a QCD$_3$-Chern-Simons theory.

We fractionalize a spinless electron operator into three fermionic partons
\begin{equation}\label{eq:SU3}
 c = f f_1 f_2.  
\end{equation}
The $\sigma^{xy}=1/3$ state is realized by a mean-field ansatz that all the fermionic partons are in a $C=1$ band (or Landau level), while the $\sigma^{xy}=-1$ state is realized by a mean-field ansatz that $f$ is in a $C=1$ band and $f_{1,2}$ are in a $C=-1$ band, respectively.
The transition between these two states can be described by tuning the Chern number of $f_{1,2}$ from $C=1$ to $C=-1$.
For each fermionic parton we can use lattice symmetries (eg. magentic translation symmetry) to protect a Chern number transition with $\Delta C=2$.
However, there is no global symmetry that can force $f_1$ and $f_2$ to change their Chern number simultaneously.
In other words, in a naive setup this phase transition cannot be a direct phase transition. 

It turns out that the emergent non-Abelian gauge symmetry can help to realize a direct phase transition.
The parton construction (Eq.~\eqref{eq:SU3}) has a maximal $SU(3)$ gauge symmetry, $(f, f_1, f_2)$ forms a $SU(3)$ fundamental.
This $SU(3)$ gauge structure can be kept in the $\sigma^{xy}=1/3$ state, while it is broken down (the maximally remaining gauge symmetry is $U(2)=SU(2)\times U(1)/Z_2$) in the $\sigma^{xy}=-1$ state.
We can enforce the emergent gauge symmetry is $U(2)$, for which $(f_1, f_2)$ is a $U(2)$ fundamental, while the $f$ parton is charged under the diagonal $U(1)$ of the $U(2)$ gauge symmetry. 
The $f_1$ and $f_2$ partons are now related by the $U(2)$ gauge rotation, hence their Chern number has to change simultaneously.  
 The critical theory is described by an effective Lagrangian:
\begin{equation}\label{eq:U2critical}
\begin{split}
\mathcal L&=\sum_{I=1}^{2}\bar{\psi}_I(i\partial \!\!\!/+\alpha \!\!\!/-m)\psi_I\\
&+\frac{1}{4\pi}(A-\mathrm{Tr}(\alpha))d(A-\mathrm{Tr}(\alpha))+\mathrm{CS}_g,
\end{split}
\end{equation}
where $\alpha$ is a $U(2)$ spin gauge field, and $\psi_I$ are $U(2)$ fundamental Dirac fermions. 
The mass term $m\sum_{I=1}^2 \bar \psi_I \psi_I$ is the tuning operator for the phase transition.
 At semi-classical regime $|m|\gg 1$, the Dirac fermion can be integrated out and we end up with
\begin{align} \label{eq:U2TQFT}
\mathcal L_{\mathrm{eff}}&=\frac{\mathrm{sgn}(m)}{4\pi}\mathrm{Tr}(\alpha d\alpha+\frac{2}{3}\alpha^3)+(2\mathrm{sgn}(m)+1)\mathrm{CS}_g \nonumber \\
& + \frac{1}{4\pi}(A-\mathrm{Tr}(\alpha))d(A-\mathrm{Tr}(\alpha)).
\end{align}
Depending the sign of $m$, one would either get a $\sigma^{xy}=1/3$ state ($m\gg 1$) or a $\sigma^{xy}=-1$ state ($m\ll -1$).
One may worry that the above $U(2)^s$ gauge theory may not describe the Abelian state. 
To work out its topological order, we need to utilize the level-rank duality. 
Let us explain it for the case of $m\gg 1$.
We start with the level rank duality, $U(N)_{-K, -K}^s\cong SU(K)_{N}$ (with $N=2, K=1$):
\begin{align}
&\frac{1}{4\pi} \mathrm{Tr}(\alpha d\alpha+\frac{2}{3}\alpha^3)-\frac{1}{2\pi} B d\mathrm{Tr}(\alpha) + \frac{2}{4\pi} BdB + 2\mathrm{CS}_g \nonumber \\
& \cong \, \mathrm{Trivial}.
\end{align}
Next we implement $S,T$ transformation $S\star T^{-3}$ on the above duality~\footnote{$T^{-3}$ corresponds to adding a level $3$ background Chern-Simons term $-\frac{3}{4\pi} B dB$, and $S$ corresponds to promoting background field $B$ to a dynamical gauge field $b$ on the both side of duality~\cite{seiberg2016duality}.}, yielding (a background term $Adb$ has been automatically added),
\begin{align}
&\frac{1}{4\pi} \mathrm{Tr}(\alpha d\alpha+\frac{2}{3}\alpha^3)+\frac{1}{2\pi} b d(A-\mathrm{Tr}(\alpha)) -\frac{1}{4\pi} bdb + 2\mathrm{CS}_g \nonumber \\
& \cong\, -\frac{3}{4\pi} b d b + \frac{1}{2\pi} A d b. \label{eq:duality_FQH}
\end{align}
Integrating out the $U(1)$ gauge field $b$ in the first Lagrangian, we eventually end up with,
\begin{align}
\mathcal L_{\mathrm{eff}}&=\frac{1}{4\pi}\mathrm{Tr}(\alpha d\alpha+\frac{2}{3}\alpha^3)+3\mathrm{CS}_g \nonumber \\
& + \frac{1}{4\pi}(A-\mathrm{Tr}(\alpha))d(A-\mathrm{Tr}(\alpha)),
\end{align}
which is exactly Eq.~\eqref{eq:U2TQFT} for $m\gg 1$.
Also the second Lagrangian of the duality in Eq.~\eqref{eq:duality_FQH} is the effective theory of $\sigma^{xy}=1/3$ state.
Therefore, we have shown that $m\gg 1$ limit gives the $\sigma^{xy}=1/3$ state. 
Similarly, one can show $m\ll -1$ gives the $\sigma^{xy}=-1$ state.

\subsection{Generalization to $U(N)$ gauge theories}

We now consider a generalization of the previous discussion.
We first fractionalize a particle, 
\begin{equation}
c = \varphi_1 \varphi_2\cdots \varphi_k f_1\cdots f_n.
\end{equation}
Here $\varphi_i$, $f_i$ are fermionic partons, and $k+n$ is odd (even) if the original particle $c$ is fermionic (bosonic).
We consider the transition between two states: 1) $\sigma^{xy}=1/(k+n)$ state, for which both $\varphi_i$ and $f_i$ are in $C=1$ bands; 
2) $\sigma^{xy}=1/(k-n)$ state, for which $\varphi_i$ is in $C=1$ bands while $f_i$ is in $C=-1$ bands.
The critical theory is,
\begin{equation}
\begin{split}\label{eq:Uncritical1}
\mathcal L&=\sum_{I=1}^{2}\bar{\psi}_I(i\partial \!\!\!/+\alpha \!\!\!/-m)\psi_I-\frac{k}{4\pi} b db +\frac{1}{2\pi} b d (A-\mathrm{Tr}\alpha).
\end{split}
\end{equation}
Here $\alpha$ is a $U(n)$ spin gauge field, $\psi_I$ is in the $U(n)$ fundamental representation. 
$b$ is a $U(1)$ gauge field, which comes from the $\varphi$ partons (one can view $\varphi_1\cdots \varphi_k$ as a $1/k$ Laughlin state).
If $k=1$, one can integrate out $b$ yielding a similar form of Eq.~\eqref{eq:U2critical}.
For $k>1$, one cannot integrate out $b$ as it represents a topological order. The tuning parameter of the transition is again the mass term of the Dirac fermions, $m\sum_{I=1}^2\bar{\psi_I}\psi_I$. At semi-classical regime $|m|\gg 1$, we can integrate out the matter fields and the resulting theory is 
\begin{equation} \label{eq:UnTQFT}
\begin{split}
\mathcal L_{\mathrm{eff}}=\mathrm{sgn}(m)[\frac{1}{4\pi}\mathrm{Tr}(\alpha d\alpha+\frac{2}{3}\alpha^3)+n\mathrm{CS}_g]\\
-\frac{k}{4\pi} b db +\frac{1}{2\pi} b d (A-\mathrm{Tr}\alpha).
\end{split}
\end{equation}
Similar as before we can use  the level-rank duality to work out the topological order described by the above theory. 
The $U(n)^s_{-1}$ theory is dual to a topologically trivial theory [\onlinecite{hsin2016level}]
\begin{equation}
    \frac{1}{4\pi}\mathrm{Tr}(\alpha d\alpha+\frac{2}{3}\alpha^3)-\frac{1}{2\pi}\mathrm{Tr}(\alpha) dB\cong -\frac{n}{4\pi}BdB-n\mathrm{CS}_g.
\end{equation}
We then implement a transformation $S\star T^{-k}$ on both sides of the duality, yielding

\begin{align}
    &\frac{1}{4\pi}\mathrm{Tr}(\alpha d\alpha+\frac{2}{3}\alpha^3)+n\mathrm{CS}_g-\frac{k}{4\pi} b db +\frac{1}{2\pi} b d (A-\mathrm{Tr}\alpha) \nonumber
    \\
    &\cong -\frac{n+k}{4\pi}bdb+\frac{1}{2\pi}bdA.
\label{eq:n+k}
\end{align}
The right hand side of Eq. (\ref{eq:n+k}) is exactly the TQFT of $\sigma^{xy}=1/(n+k)$ Abelian fractional quantum Hall/Chern insulator state that appears when $m\gg 1$. 
The same analysis can also be implemented to the $m\ll -1$ case, yielding the $\sigma^{xy}=1/(n-k)$  Abelian fractional quantum Hall/Chern insulator state. 

We remark that the above discussion carries over for $k=0$ with a slight modification that the gauge field changes from $U(N)$ to $SU(N)$.

Similarly, we use another construction, 
\begin{equation}
\label{eq:abelianparton}
c = \varphi\varphi_1 \varphi_2\cdots \varphi_k f_1\cdots f_n.
\end{equation}
Here $\varphi$, $\varphi_i$, $f_i$ are all fermionic partons, and $k+n$ is odd (even) if the physical particle $c$ is bosonic (fermionic). We consider the transition between two states, namely 1) $\sigma^{xy}=p/((k+n)p+1)$ state: $\varphi$ in a $C=p$ band, $\varphi_i$ and $f_i$ in $C=1$ bands; 2) $\sigma^{xy}=p/((k-n)p+1)$ state: $\varphi$ in a $C=p$ band, $\varphi_i$ in $C=1$ bands while $f_i$ in $C=-1$ bands. The critical theory has again two flavors of Dirac fermions, interacting with a $U(n)$ spin gauge field,
\begin{align}
 \label{eq:Uncritical2}
 \mathcal L&=\sum_{I=1}^{2}\bar{\psi}_I(i\partial \!\!\!/+\alpha \!\!\!/-m)\psi_I-\frac{k}{4\pi} b db \nonumber\\ 
&+\frac{1}{2\pi} b d (\beta-\mathrm{Tr}\alpha)+\frac{p}{4\pi}(A-\beta) d(A-\beta)+p\mathrm{CS}_g.
\end{align}
Here $\alpha$ is a  $U(n)$ spin gauge field, $\psi_I$ is in the $U(n)$ fundamental representation. $b$, $\beta$ are $U(1)$ gauge field and spin gauge field, respectively. 
The first three terms in Eq.~\eqref{eq:Uncritical2} are similar to the critical theory Eq.~\eqref{eq:Uncritical1}. 
In the third term we have a dynamical $U(1)$ spin gauge field $\beta$, which corresponds to the $U(1)$ gauge symmetry coming from the relative phase rotation between $\varphi$ and $\varphi_1 \varphi_2\cdots \varphi_k f_1\cdots f_n$ in the parton construction Eq. (\ref{eq:abelianparton}). 
Similar as before one can show that at semi-classical regime $|m|\gg 1$, the effective theory (after applying the level-rank duality) is
\begin{equation}
\label{eq:fluxtransition}
\begin{split}
        \mathcal L_{\mathrm{eff}}= -\frac{\mathrm{sgn}(m)n+k}{4\pi}bdb+\frac{1}{2\pi}b da
        \\
       +\frac{p}{4\pi}(A-\beta) d(A-\beta)+p\mathrm{CS}_g,
    \end{split}
\end{equation}
which is exactly the effective theory of Abelian composite fermion states with 1) $\sigma^{xy}=p/((k+n)p+1)$ state when  $m\gg 1$; and 2) $\sigma^{xy}=p/((k-n)p+1)$ state when $m\ll -1$.

\section{\label{sec:parton} Parton Constructions for Non-Abelian States}

In the previous section we showed that a direct phase transition between two FQHs/FCIs in different Jain sequences [\onlinecite{ezawa2008quantum}] are necessarily described by a QCD$_3$-Chern-Simons theory. 
In the rest of this paper we study phase transitions involving non-Abelian FQHs/FCIs , i.e., states that have anyons with non-Abelian statistics. 
We focus on non-Abelian states  that are close to the Moore-Read ``Pfaffian" state~\cite{moore1991nonabelions}, including Pfaffian, anti-Pfaffian, PH-Pfaffian, bosonic Pfaffian states, Ising topological order, etc.
Some of them might be realized in the half filled Landau level~\cite{willett1987observation,dolev2008observation,banerjee2018observation,zibrov2018even} or frustrated spin systems~\cite{kitaev2006anyons}.
The transitions involving these states can be naturally captured by QCD$_3$-Chern-Simons theories with different gauge groups. 
For this purpose, we first discuss the parton constructions for these states. 
Then in Sec. \ref{sec:tran}, the phase transitions will be analyzed.

\subsection{$USp(4)$ parton construction}

Many of the non-Abelian topological orders we discuss below contain a $USp(4)^s$ spin gauge field in their TQFT descriptions. 
In this subsection we will first introduce a parton construction that has emergent $USp(4)$ gauge symmetry~\cite{wen1999projective,Barkeshli_USpN}. 
This $USp(4)$ parton construction is similar to the familiar $SU(2)$ parton construction ($\vec S=f^\dag \vec \sigma f)$, and it is compatible with a global $SO(3)$ spin rotation symmetry. 

We consider a spin system (or equivalently a bosonic system), and fractionalize the spin operator into fermionic partons,
\begin{equation}
S^-=S^x-iS^y=(\psi_1\psi_4-\psi_3\psi_2),
\label{eq:bpfparton}
\end{equation}
which maximally has a local $USp(4)$ gauge invariance, with $\Psi=(\psi_1,\,\psi_2,\,\psi_4,\,\psi_3)^T$ being a $USp(4)$ vector (fundamental). 
The local constraint is,
\begin{equation}
n_1=n_4, \quad n_2=n_3,
\end{equation}
on each site.
$S^-$ can be expressed as
\begin{equation}
S^-=\frac{1}{2}\Psi^T
\begin{pmatrix} 
0&0 & 1&0 \\ 
0 & 0 &0 &1 \\
-1&0&0&0\\
0&-1&0&0
\end{pmatrix}
\Psi,
\end{equation}
which is invariant under a local $USp(4)$ transformation $\Psi\to L\Psi$. Note that the local Hilbert space has spin $S=1$ since $(S^-)^2\ne0$ but $(S^-)^3=0$. The $USp(4)$ gauge invariance and the physical spin rotation symmetry can be easily seen by introducing a $4\times 2$ matrix $X$,
\begin{equation}
X=
\begin{pmatrix} 
\psi_1^\dagger & -\psi_4 \\ 
 \psi_2^\dagger  &-\psi_3 \\
\psi_4^\dagger&\psi_1\\
\psi_3^\dagger&\psi_2
\end{pmatrix},
\end{equation}
which satisfies a reality condition
\begin{equation}
X^*=\mu^2X\sigma^2.
\label{eq:real}
\end{equation}
Here $\mu^2$ and $\sigma^2$ are Pauli matrices acting from left and right, respectively:
\begin{equation}
\mu^2=
\begin{pmatrix} 
0&0 & -i&0 \\ 
0 & 0 &0 &-i \\
i&0&0&0\\
0&i&0&0
\end{pmatrix},
\end{equation}
\begin{equation}
\sigma^2=
\begin{pmatrix} 
0&-i \\ 
i&0
\end{pmatrix}.
\end{equation}

The physical spin operators can  be expressed as
\begin{equation}
S^i=-\frac{1}{4}\mathrm{Tr}(X^\dagger X\sigma^i),
\end{equation}
in which spin rotation symmetry acts on $X$ from right
\begin{equation}
X\to X(U^s)^\dagger.
\end{equation}
It is also easy to see that the spin operator is invariant under a $USp(4)$ gauge transformation, acting from left,
\begin{equation}
X\to LX.
\end{equation}
In general, any elements in $U(4)$ leave spin operators invariant, but the reality condition Eq.(\ref{eq:real}) requires $L$ lie in $USp(4)$, namely $L^T\mu^2L=\mu^2$. Eq.(\ref{eq:real}) on the other hand does not impose further constraints on the global spin rotation symmetry. 
Note that the global symmetry $SU(2)$ and the gauge group $USp(4)$ share a common center $-1$. Thus the physical global symmetry is actually $SU(2)/Z_2=SO(3)$. 

This parton construction can also be understood intuitively by noting that the maximal faithful symmetry of four complex fermions $\psi_i$ $(i=1,2,3,4)$ is $O(8)$.
After gauging a $USp(4)\sim SO(5)$ subgroup, an $SO(3)$ subgroup is left as the global spin rotation symmetry.

We also remark that, explicitly the parton construction is,

\begin{equation}
S^z=\frac{1}{2}(\psi_1^\dagger\psi_1+\psi_2^\dagger\psi_2-\psi_3\psi_3^\dagger-\psi_4\psi_4^\dagger),
\end{equation}   
\begin{equation}
S^+=S^x+iS^y=-\psi_1^\dagger\psi_4^\dagger+\psi_3^\dagger\psi_2^\dagger,
\end{equation}  
\begin{equation}
S^-=S^x-iS^y=\psi_1\psi_4-\psi_3\psi_2.
\end{equation}  
It is straightforward to check that they satisfy the standard commutation relation
\begin{equation}
[S^i,S^j]=i\varepsilon^{ijk}S^k.
\end{equation} 

\subsection{Partons for Pfaffian-like states}

Using similar results from level-rank duality, we are able to find the parton construction of different non-Abelian states, which are summarized in Table \ref{tab:nonabelian}. 
Some of these constructions have been discussed in previous papers~\cite{wen1991non,wen1999projective,Barkeshli_USpN,zou2018field,balram2018parton}.

\begin{table*}[ht]
\caption{Summary of Parton Constructions of Pfaffian-like states.~\footnote{ When $\kappa_{xy}$ is concerned in the literature of the $\nu=5/2$ quantum Hall state, an extra $\kappa_{xy}=2$ from the filled first Landau level is usually included.} } \label{tab:nonabelian}
\setlength{\tabcolsep}{0.2cm}
\renewcommand{\arraystretch}{1.4}
\centering
\begin{tabular}{ccccc}
\hline  
State &$\kappa_{xy}$  & Gauge field & Spin gauge field & Parton  \\
\hline
Ising topological order & $1/2$ & $U(2)_{2,-2}$ & $\frac{U(2)_{-2,-2}^s\times U(1)_{-4}}{Z_2}$  & $S^+=\phi^\dagger f_1^\dagger f_2^\dagger$  \\
\hline
Bosonic $U(2)_2$ state & $5/2$  &  $U(2)_{2,2}$  &  $SU(2)_{-2}^s$ &    $\vec{S}= f^\dag \vec \sigma f/2$
\\
\hline
Bosonic Pfaffian & $3/2$ & $SU(2)_2$ & $U(2)^s_{-2,-2}\cong USp(4)^s_{-1}$ & $S^-=(\psi_1\psi_4-\psi_3\psi_2)$ \\
\hline
Wen's $(221)$-parton state & $5/2$  & $U(2)_{2,4}$ & $U(2)_{-2,-4}^s$ &  $c=f_1f_2\psi$ 
\\
\hline
Pfaffian & $3/2$ & $\frac{\mathrm{Ising}\times U(1)_8}{Z_2}$ & $\frac{U(2)^s_{-2,-2}\times U(1)_{-8}}{Z_2}\times U(1)_1\cong \frac{USp(4)^s_{-1}\times U(1)_{-8}}{Z_2}\times U(1)_1$ & $c=\frac{1}{\sqrt{2}}(\psi_1\psi_4-\psi_3\psi_2)\psi$\\
\hline
Anti-Pfaffian & $-1/2$ & $U(2)_{-2,4}$ & $\frac{U(2)_{2,2}^s\times U(1)_{8}}{Z_2}$ & $c=f_1f_2 \psi$ \\
\hline
PH-Pfaffian & $1/2$ & $\frac{\overline{\mathrm{Ising}}\times U(1)_8}{Z_2}$ & $\frac{\frac{USp(4)_1^s\times U(1)_{4}}{Z_2}\times U(1)_8}{Z_2}$ & $c=\frac{1}{\sqrt{2}}(\psi_1\psi_4-\psi_3\psi_2)\psi$ \\
\hline
\end{tabular}
\end{table*}

\subsubsection{Bosonic $U(2)_2$ state}
As discussed above, the bosonic $U(2)_2$ topological order can be realized in spin/bosonic systems. 
Without loss of generality, we take the spin-1/2 system as an example. We start with the standard $SU(2)$ parton construction
\begin{equation}
    \vec{S}= f^\dag \vec \sigma f/2, 
\end{equation}
where $f^\dag=(f_\uparrow^\dag, f_\downarrow^\dag)$. This parton construction has a local $SU(2)$ gauge structure. The $U(2)_2$ topological order is realized by putting both $f_{\uparrow,\downarrow}$ in $C=2$ bands~\cite{wen1991non,wen1999projective}. 
After integrating out the partons, one ends up with an $SU(2)^s_{-2}$ Chern-Simons theory
\begin{equation}
    \frac{2}{4\pi}\mathrm{Tr}[(\alpha+\frac{A}{2}\mathbf{1_2}) d (\alpha+\frac{A}{2}\mathbf{1_2})+\frac{2}{3}(\alpha+\frac{A}{2}\mathbf{1_2})^3]+4\mathrm{CS}_g.
\end{equation}
Here $\alpha$ is a spin $SU(2)$ gauge field. $A$ is the $U(1)$ background field coupled to the global $S^z$ spin rotation. Using a level/rank duality $SU(2)^s_{-2}\cong U(2)_{2,2}$, the description with normal gauge field can be obtained:
\begin{equation}
    -\frac{2}{4\pi}\mathrm{Tr}(ada+\frac{2}{3}a^3)+\frac{1}{2\pi}Ad(\mathrm{Tr}a).
\end{equation}
Here $a$ is a $U(2)$ gauge field.

\subsubsection{Ising topological order}

The Ising topological order corresponds to the $\nu=1$ state in Kitaev's 16-fold-way \cite{kitaev2006anyons}. It can be realized on a model of localized spin-$1/2$ particles on a honeycomb lattice, with a designed spin interactions. The TQFT description of Ising topological order in terms of spin gauge fields and parton construction were derived in detail in Ref. \cite{zou2018field}, here we briefly summarize the results.

The Ising topological order is described by $U(2)_{2,-2}\cong \frac{SU(2)_2\times U(1)_{-4}}{Z_2}$, which was shown to be equivalent to the following topological order
\begin{equation}
    \frac{U(2)_{-2,-2}^s\times U(1)_{-4}}{Z_2}
    \label{eq:Isingspin}
\end{equation}
with Lagrangian
\begin{equation}
    \mathcal L_{Ising}=\frac{2}{4\pi}\mathrm{Tr}(\chi d\chi+\frac{2}{3}\chi^3)+\frac{1}{2\pi}a d\mathrm{Tr}\chi +\frac{2}{4\pi}a da+ 4\mathrm{CS}_g,
    \label{eq:Ising}
\end{equation}
where $\chi$ is a spin $U(2)$ gauge field and $a$ is a $U(1)$ gauge field. The parton construction is achieved by fractionalizing the physical spin operator as
\begin{equation}
    S^-=\phi f_1 f_2.
\end{equation}
This parton construction has a U(2) gauge invariance: $(f_1,f_2)^T$ is in the $U(2)$ fundamental representation, and it is interacting with a $U(2)$ spin gauge field $\chi$; $\phi$ carries charge under the $U(1)$ diagonal part of $\chi$.\par

To get the Ising topological order, we put the bosonic parton $\phi$ into a $\nu=-\frac{1}{2}$ Laughlin state, and put the fermionic partons $f_i$ into a topological band with Chern number $C=2$. After integrating out the gapped partons, we end up with the TQFT Eq. (\ref{eq:Ising}).

\subsubsection{Bosonic Pfaffian state}

The bosonic Pfaffian state is believed to be realized by bosons in the first Landau level at filling factor $\nu=1$, and it can be understood as the $p+ip$ pairing state of composite fermions~\cite{nayak,read2000paired}. 
At long wavelength the bosonic Pfaffian state can be described by a $SU(2)_2$ Chern-Simons gauge theory~\cite{nayak},
\begin{equation}\label{eq:SU2_2}
-\frac{2}{4\pi}\mathrm{Tr}(bdb+\frac{2}{3}b^3)+\frac{1}{4\pi}BdB,
\end{equation}
in which $b$ is a $SU(2)$ gauge field, $B$ is a background probe field. 

It is known that the bosonic Pfaffian state can also be realized in a spin$-1$ system preserving the $SO(3)$ spin rotation symmetry, so sometimes it is also refered as a non-Abelian chiral spin liquid \cite{greiter2009non}.
To construct the bosonic Pfaffian state, we use Eq.(\ref{eq:bpfparton}) to fractionalize the spin-1 operator into fermionic partons~\footnote{For the bosonic system, one can simply replace spin operators by bosonic annihilation/creation operators.}.
The bosonic Pfaffian state is realized by having $\psi_i$ in a $C=1$ band~\cite{wen1999projective,Barkeshli_USpN}, yielding a $USp(4)_{-1}^s$ Chern-Simons theory ($\beta$ is a $USp(4)$ spin gauge field),
\begin{equation}
\frac{1}{4\pi} \Tr(\beta d \beta +\frac{2}{3} \beta^3) + \frac{1}{4\pi} B d B + 4\gcs,
\label{eq:USp4_-1}
\end{equation}
which is dual to $SU(2)_{2}$ in Eq.~\eqref{eq:SU2_2} through the level-rank duality.
As a self-consistent check, one can find that after integrating the $USp(4)$ gauge field in the above $USp(4)$ theory has the same thermal Hall response as of the bosonic Pfaffian ($\kappa_{xy}=3/2$).

We can break the $USp(4)$ gauge symmetry down to a $U(2)$ gauge symmetry, and the $USp(4)_{-1}^s$ Chern-Simons theory will become the $U(2)_{-2,-2}^s$ theory. 
Interestingly, $U(2)_{-2,-2}^s$ is dual to $USp(4)_{-1}^s$ (similar to $SU(2)^s_{-1}\cong U(1)^s_{-2})$), hence this gives us an alternative way to describe bosonic Pfaffian as well as their transitions.

\subsubsection{Pfaffian state}
The fermionic Pfaffian state~\cite{moore1991nonabelions} is described by the following TQFT [\onlinecite{lian2018theory}]
\begin{equation}
\frac{\mathrm{Ising}\times U(1)_8}{Z_2}\cong \frac{U(2)_{2,-2}\times U(1)_8}{Z_2},
\end{equation}
with the Lagrangian [\onlinecite{seiberg2016gapped}]
\begin{align}
\mathcal L_{pf}=&-\frac{2}{4\pi}\mathrm{Tr}(bdb+\frac{2}{3}b^3)+\frac{2}{4\pi}\mathrm{Tr}bd\mathrm{Tr}b \nonumber\\
&-\frac{1}{2\pi}(\mathrm{Tr}b)dc-\frac{1}{4\pi}cdc+\frac{1}{2\pi}Adc.
\label{eq:pfbosonic}
\end{align}
Here $b$ is a $U(2)$ gauge field, $c$ is a $U(1)$ gauge field. 
It is shown in Appendix \ref{app:duality} that its dual TQFT description in terms of spin gauge fields are 
\begin{equation}
\begin{split}
\mathcal L_{pf}&=\frac{1}{4\pi}\mathrm{Tr}(\beta d \beta+\frac{2}{3}\beta^3)+\frac{1}{4\pi}a d a\\
&-\frac{1}{4\pi}b d b+\frac{1}{2\pi}(A+a)db+4\mathrm{CS}_g,
\label{eq:pffermionusp}
\end{split}
\end{equation}
in which $\beta$ is a  $USp(4)$ spin gauge field and $a$, $b$ are both $U(1)$ gauge fields. 
The chiral central charge of the theory is $-\frac{5}{2}-1+1+4=\frac{3}{2}$ ($-\frac{5}{2}$ comes from $USp(4)_{-1}$ part, $-1+1$ comes from the two $U(1)$ parts and $4$ comes from $4\mathrm{CS}_g$), which is identical to that of the Pfaffian state. 

This TQFT motivates a parton construction for electrons \cite{wen1999projective, Barkeshli_USpN}
\begin{equation}
c=\frac{1}{\sqrt{2}}(\psi_1\psi_4-\psi_3\psi_2)\psi,
\label{eq:pfaparton}
\end{equation}
which has a local $\frac{USp(4)\times U(1)}{Z_2}$ gauge redundancy. Roughly speaking, $\psi_i$ ($i=1,2,3,4$) furnish a fundamental representation of $USp(4)$ subgroup and we will put them on $C=1$ Chern bands. We also put $\psi$ on a $C=1$ band. Besides the dynamical $USp(4)$ spin gauge field $\beta$, $\psi_i$'s also couple to $\frac{a}{2}\mathbf{1_4}$, where $a$ is a  $U(1)$ gauge field which comes from the relative phase rotation between $(\psi_1\psi_4-\psi_3\psi_2)$ and $\psi$. And $\psi$ has charge 1 under $A$, the external electromagnetic field. Gauge invariance imposes a constraint on the density of the partons:
\begin{equation}
n_1=n_2=n_3=n_4=\frac{n_\psi}{2}=\frac{n}{2}.
\end{equation}
\label{eq:pfTQFT2}
Integrating out the fermions leads to
\begin{align}
\label{eq:Integratepartonpf}
&\frac{1}{4\pi}\mathrm{Tr}\left[(\beta+\frac{a}{2}\mathbf{1_4})d(\beta+\frac{a}{2}\mathbf{1_4})+\frac{2}{3}(\beta+\frac{a}{2}\mathbf{1_4})^3\right]
 \nonumber
\\&
+\frac{1}{2\pi}(a+A) d b-\frac{1}{4\pi}b db+4\mathrm{CS}_g,
\end{align}
in which the $b$ field describes the $\nu=1$ integer Quantum Hall state of $\psi$. We thus recover the TQFT of fermionic Pfaffian state Eq.~\eqref{eq:pffermionusp}.

The fractional $U(1)$ gauge charge of the $\psi_i$ partons looks ill-defined. Actually it reflects the fact that the center element of $USp(4)$ should be identified with $\pi$ flux of the $U(1)$ gauge symmetry, and the precise gauge structure of Pfaffian state is $\frac{USp(4)^s_{-1}\times U(1)_{-8}}{Z_2}\times U(1)_1$. Detailed discussion is shown in Appendix \ref{app:duality}. One more subtlety is that here an ordinary gauge field $a$, instead of a spin gauge field, is coupled to a fermion $\psi$. From the parton construction Eq. (\ref{eq:pfaparton}), one can see this field couples to both a bosonic part, i.e. $\sim(\psi_1\psi_4-\psi_3\psi_2)$, and a fermionic part, i.e. $\psi$. These two viewpoints differ by a local fermion, namely the physical electron, which has no influence on the topological order in electronic systems. If we interpret the electromagnetic field $A$ as a $\mathrm{spin}_c$ connection, the choice we adopt here can be consistently put on a $\mathrm{spin}_c$ manifold, since $(a+A)$, to which $\psi$ couples, is a $\mathrm{spin}_c$ connection.

\subsubsection{Anti-Pfaffian state}
The anti-Pfaffian state~\cite{LeeAPf,LevinAPf} is described by
\begin{equation}
\begin{split}
U(2)_{-2,4}&\cong \frac{SU(2)_{-2}\times U(1)_8}{Z_2}\\
&\cong \frac{2}{4\pi}\mathrm{Tr}(bdb+\frac{2}{3}b^3)-\frac{3}{4\pi}\mathrm{Tr}bd\mathrm{Tr}b+\frac{1}{2\pi}\mathrm{Tr}bdA,
\end{split}
\end{equation}
where $b$ is a $U(2)$ gauge field which is coupled to a bosonic matter, $A$ is the background electromagnetic field. The dual spin gauge field description can be found by the similar method as that described for Pfaffian state and the result is (see Appendix~\ref{app:duality} for details)
\begin{equation}
\label{eq:apffermionic}
\mathcal L_{apf}=-\frac{2}{4\pi}\mathrm{Tr}(\chi d\chi+\frac{2}{3}\chi^3)-\frac{3}{4\pi}cdc+\frac{1}{2\pi}cd(\mathrm{Tr}\chi+A)-4\mathrm{CS}_g,
\end{equation}
which is roughly a \textbf{$U(2)^s\times U(1)$} theory.
The chiral central charge of the TQFT in Eq. (\ref{eq:apffermionic}) can then be easily obtained: $\frac{5}{2}+1-4=-\frac{1}{2}$,  consistent the Anti-Pfaffian state. 

This TQFT motivates a parton construction for Anti-Pfaffian state [\onlinecite{balram2018parton}],
\begin{equation}
c=f_1f_2 \psi,
\label{eq:antiparton}
\end{equation} 
where we enforce the emergent gauge symmetry is $\frac{SU(2)\times U(1)}{Z_2}\cong U(2)$, and $(f_1,\,f_2)^T$ is a $U(2)$ fundamental coupled to a dynamical $U(2)$ spin gauge field $\chi$, while $\psi$ is charged under the diagonal $U(1)$ of the $U(2)$ gauge field. 
We put $f_{1,2}$ in $C=-2$ bands and assume $\psi$ forms a $\nu=\frac{1}{3}$ Laughlin state. Integrating out the fermion fields gives rise to Eq.~\eqref{eq:apffermionic}.

\subsubsection{PH-Pfaffian state}

The PH-Pfaffian state~\cite{FidkowskiPHPf} can be described by the following Chern-Simons theory in terms of gauge field:
\begin{equation}
\frac{\overline{\mathrm{Ising}}\times U(1)_8}{Z_2}\cong \frac{U(2)_{-2,2}\times U(1)_{8}}{Z_2},
\end{equation}
with Lagrangian
\begin{align}
\mathcal L_{phpf}=&\frac{2}{4\pi}\mathrm{Tr}(bdb+\frac{2}{3}b^3)-\frac{2}{4\pi}\mathrm{Tr}bd\mathrm{Tr}b+\frac{1}{2\pi}\mathrm{Tr}bdc\nonumber\\&-\frac{3}{4\pi}cdc-\frac{1}{2\pi}Adc.
\end{align}
The dual description in terms of spin gauge field is 
\begin{align}
\mathcal L_{phpf}=&-\frac{1}{4\pi}\mathrm{Tr}(\beta d\beta+\frac{2}{3}\beta^3)-\frac{1}{4\pi}b d b-\frac{1}{2\pi}b da \nonumber\\
&-\frac{3}{4\pi}a d a+\frac{1}{2\pi}Ada-4\mathrm{CS}_g
\label{eq:phfermionic}
\end{align}
where $\beta$ is a  $USp(4)$ spin gauge field, $b$ and $a$ are $U(1)$ gauge fields. 
The duality is derived in Appendix \ref{app:duality}. 
We can do a self-consistent check by examining the gravitational response. 
Integrating out the gauge fields yields a gravitational Chern Simons term $\frac{9}{2}\mathrm{CS}_g$. Combined with the last term in Eq.~\eqref{eq:phfermionic}, the total gravitational response is $\frac{1}{2}\mathrm{CS}_g$, which is identical to that of the PH-Pfaffian state. The parton construction is straightforward,
\begin{equation}
c=\frac{1}{\sqrt{2}}(\psi_1\psi_4-\psi_3\psi_2)\psi.
\label{eq:PHparton}
\end{equation}
Similar to the case in Pfaffian state, this parton construction also has a $\frac{USp(4)^s\times U(1)}{Z_2}$ local gauge redundancy. We put $\psi_i,\, i=1,2,3,4$, which are in fundamental representation of $USp(4)$ group and carry charge $1/2$ under the $U(1)$ gauge field $b$, in $C=-1$ bands. The $\psi$ parton, which carries charge 1 under both the  $U(1)$ gauge field $b$ and external electromagnetic field $A$, forms a $\nu=\frac{1}{3}$ Laughlin state. After integrating out the fermionic matters, one reproduces the TQFT for PH-Pfaffian state, Eq. (\ref{eq:phfermionic}).
A detailed analysis on normalization conditions for gauge charge and gauge flux shows that this Lagrangian can be written as (see Appendix. \ref{app:duality}) \begin{equation}
       \frac{\frac{USp(4)_1^s\times U(1)_{4}}{Z_2}\times U(1)_8}{Z_2}.
       \label{eq:PHpffermion}
\end{equation}  

\subsubsection{Wen's $(221)$-parton state}
Wen's $(221)$-parton state~\cite{wen1991non,wen1999projective} was recently proposed as a candidate state of the half filled higher Landau levels of graphene~\cite{kim2019even}. 
Its parton construction is
\begin{equation}
    c=f_1f_2\psi
\end{equation}
where $f_1$, $f_2$ and $\psi$ are all fermionic partons. Note that the construction is the same as that for Anti-Pfaffian state, Eq.(\ref{eq:antiparton}), which has a $U(2)\cong SU(2)\times U(1)/Z_2$ gauge redundancy. The $U(2)$ doublet $(f_1\,f_2)$ is coupled to a dynamical $U(2)^s$ gauge field $\chi$ while $\psi$ is charged under $\mathrm{Tr}\chi$. We put $(f_1\,f_2)$ in $C=2$ bands and $\psi$ in a $C=1$ band. After integrating out the partons, one ends up with an effective theory 
\begin{equation}
    \begin{split}
        L_{221}=&\frac{2}{4\pi}\mathrm{Tr}(\chi d\chi+\frac{2}{3}\chi^3)
        \\&+\frac{1}{4\pi}(\mathrm{Tr}\chi+A) d (\mathrm{Tr}\chi+A)+5\mathrm{CS}_g,
    \end{split}
\end{equation}
which is the $U(2)_{-2,-4}^s$  Chern-Simons theory.

\section{\label{sec:tran} Phase transitions from the Pfaffian-like states}

Following the philosophy used in the discussion about transitions between Abelian FCIs/FQHs in Section \ref{sec:abelian2}, we can also describe  phase transitions involving these Pfaffian-like states. 
We summarize the key ideas: (1) Phase transitions in general can be understood as Chern number changing transitions of the underlying parton insulators, which are mediated by $|\Delta C|$ Dirac cones. At the critical point, $N_f=|\Delta C|$ flavors of Dirac fermions interact with non-Abelian gauge fields. (2) Naively only $|\Delta C|=1$ transitions are generic without fine tuning, however one can use symmetries, in particular the magnetic translation symmetry, to enforce $|\Delta C|>1$ without fine-tuning. 
Specifically, we can design an external magnetic field and electron density such that the partons see a fractional effective flux. Thus higher Chern number changing transitions can be protected by the magnetic algebra. 
We focus on those transitions between the Pfaffian-like states and Abelian states, which is more realistic in experimental settings.

\subsection{Phase transitions out of the bosonic Pfaffian}

\begin{table*}[htb]
\caption{Transitions out of the bosonic Pfaffian can be described by the parton construction $S^-=(\psi_1\psi_4-\psi_3\psi_2)$.} \label{tab:bosonicpfaff}
\setlength{\tabcolsep}{0.2cm}
\renewcommand{\arraystretch}{1.4}
\centering
\begin{tabular}{cccccc}
\hline\hline
Phase & $(\psi_1,\psi_2)$ & $(\psi_4,\psi_3)$ & Topological order & $\sigma^{xy}$ & $\kappa^{xy}$ \\ \hline 
Bosonic Pfaffian & $C_1=1$ & $C_2=1$ & $USp(4)_{-1}^s\cong U(2)_{-2}^s\cong SU(2)_2$ & $1$ & $3/2$ \\ \hline
Mott insulator & $C_1=0$ & $C_2=0$ & $USp(4)_{0}^s$ & $0$ & $0$ \\ \hline
Reversed Bosonic Pfaffian & $C_1=-1$ & $C_2=-1$ & $USp(4)_{1}^s\cong U(2)_{2}^s\cong SU(2)_{-2}$ & $-1$ & $-3/2$ \\ \hline
Mott insulator & $C_1=0$ & $C_2=1$ & $U(2)_{-1}^s\cong\,$Trivial & $0$ & $0$ \\ \hline
Charge-2 superfluid & $C_1=-1$ & $C_2=1$ & $U(2)^s_0$ & & \\ \hline
Bosonic integer quantum Hall (SPT) & $C_1=-2$ & $C_2=1$ & $U(2)_{1}^s\cong\,$Trivial & $4$ & $0$ \\ \hline 
Reversed Bosonic Pfaffian stacked with SPT & $C_1=-3$ & $C_2=1$ & $U(2)_2^s\cong SU(2)_{-2}$ & 3 & $-3/2$ \\ \hline\hline

\end{tabular}
\end{table*}

We will first discuss phase transitions out of the bosonic Pfaffian state.
From this example it will be clear about how to understand in general the phase transitions involving other similar Pfaffian like states. 

As we discussed above the bosonic Pfaffian state can be constructed using the $USp(4)$ parton construction, $S^-=(\psi_1\psi_4-\psi_3\psi_2)$, with each fermionic parton ($\psi_{1,\cdots,4}$) on a $C=1$ Chern band.
When the fermionic partons have a different Chern number $C$, a state described by the $USp(4)_{-C}^s$ TQFT is realized.
For example, if $C=0$ we have a topologically trivial Mott insulator, while if $C=-1$ we have a time-reversal reversed bosonic Pfaffian. 
The phase transitions between the bosonic Pfaffian and these states is described by the Chern number changing transitions of the fermionic partons.
Since the fermionic partons are coupled to a $USp(4)$ spin gauge field, the critical theory of the phase transitions are
\begin{itemize}
    \item $N_f=|\Delta C|=|C-1|$ flavor of Dirac fermions coupled to a $USp(4)_{-(C+1)/2}^s$ Chern-Simons field.
\end{itemize}

It is worth noting that there is a level-rank duality between the $USp(4)_{-1}^s$ and $U(2)_{-2}^s$ Chern-Simons theory. 
Therefore, the bosonic Pfaffian state can be equivalently described by a $U(2)^s$ gauge theory. 
For the $U(2)$ parton construction, we can simply higgs the $USp(4)^s$ gauge symmetry down to the $U(2)^s$ gauge symmetry, for which $(\psi_1,\psi_2)$ and $(\psi_4,\psi_3)$ are the $U(2)$ fundamental and anti-fundamental, respectively. 
This structure is manifest if one writes the spin operator in matrix form
\begin{equation}
\label{eq:bpfu2parton1}
S^-=
\begin{pmatrix} \psi_4 & \psi_3  \end{pmatrix}
\begin{pmatrix} \psi_1 \\ \psi_2 \end{pmatrix},
\end{equation} 
and the $U(2)$ gauge transformation is simply
\begin{equation}
\label{eq:bpfu2parton}
\begin{pmatrix} \psi_1 \\ \psi_2 \end{pmatrix}\to W\begin{pmatrix} \psi_1 \\ \psi_2 \end{pmatrix},\,
\begin{pmatrix} \psi_4 \\ \psi_3 \end{pmatrix}\to W^*\begin{pmatrix} \psi_4 \\ \psi_3 \end{pmatrix},
\end{equation}
where $W$ is an $U(2)$ matrix in fundamental representation. 

By assigning Chern number $C_1$ and $C_2$ to $(\psi_1,\psi_2)$ and $(\psi_4,\psi_3)$\footnote{We remark that when $C_1\neq C_2$, the global $SO(3)$  spin rotation symmetry cannot be preserved.}, we will get a state described by the $U(2)^s_{-(C_1+C_2)}$ Chern-Simons theory, whose Lagrangian is
\begin{align}\label{eq:U2eff}
\mathcal{L}=&\frac{C_1+C_2}{4\pi} \textrm{Tr}(\alpha d \alpha + \frac{2}{3}\alpha^3)+\frac{1}{2\pi} \frac{C_1-C_2}{2} A\, d\, \textrm{Tr}(\alpha) \nonumber \\ & +\frac{1}{4\pi} \frac{C_1+C_2}{2} A d A + 2(C_1+C_2)\, \gcs.
\end{align}
Here $\alpha$ is a $U(2)$ spin gauge field and $A$ is a probe field coupled to the $S^z$ spin rotation. 
To get the response (i.e. Hall conductance $\sigma^{xy}$ and thermal Hall conductance $\kappa^{xy}$) of the state, we can
further integrating out the $U(2)$ gauge field $\alpha$,
\begin{equation}
\begin{split}
\mathcal{L}_{\textrm{res}}=&\frac{1}{4\pi} \frac{2C_1C_2}{C_1+C_2} A d A+ [2(C_1+C_2)
\\
&-\frac{3(C_1+C_2)}{2+|C_1+C_2|}-{\mathrm{sgn}(C_1+C_2)}]\gcs.
\end{split}
\end{equation}
It is worth noting that when $C_2=-C_1$, the effective theory Eq.~\eqref{eq:U2eff} reduces to $\frac{C_1}{2\pi}  A\, d\, \textrm{Tr}(\alpha)$, which describes a superfluid with charge-2$|C_1|$ condensation~\footnote{Due to the absense of Chern-Simons therm, the $U(2)$ gauge field will confine, and the monopole of the diagonal $U(1)$ field $\textrm{Tr}(\alpha)/2$ will condense. On the other hand, due to the mutual Chern-Simons term $\frac{C_1}{2\pi}  A\, d\, \textrm{Tr}(\alpha)$, the monopole will carry $q=2|C_1|$ charge of the physical particle. 
Therefore, we end up with a charge-2$|C_1|$ superfluid.}. 

The phase transitions between different states are again described by the Chern number changing transitions of partons. 
However, we shall emphasize that there is no (gauge or global) symmetry to relate $C_1$ with $C_2$, hence a direct phase transition can only have $C_1$ or $C_2$ changing, with another one fixed. 
Table. \ref{tab:bosonicpfaff} summarizes several phases obtained by varying $C_1$ with $C_2$ fixed.
The phase transitions between the bosonic Pfaffian and these states are described by the critical theory:
\begin{itemize}
\item $N_f=|\Delta C_1|=|C_1-1|$ flavor of Dirac fermions coupled to a $U(2)_{-(C_1+3)/2}^s$ Chern-Simons gauge field.
\end{itemize}

One may note that there are two different critical theories for the phase transition between the bosonic Pfaffian and Mott insulator, 1) $N_f=1$ flavor of Dirac fermions coupled to a $USp(4)^s_{-1/2}$ Chern-Simons field; 2) $N_f=1$ flavor of Dirac fermions coupled to a $U(2)^s_{-3/2}$ Chern-Simons gauge field.
Indeed these two critical theories may be dual to each other as discussed in Ref.~\cite{hsin2016level,AharonyUSpduality}.
If this duality conjecture is correct, it means that the second theory will have an emergent  $SO(3)$ global symmetry rather than the naive $U(1)$ global symmetry.

\subsection{Phase transitions of other Pfaffian-like states}

\begin{table*}[htb]
\caption{Transition out of $U(2)_2$ bosonic topological order can be described by the parton construction $S^-=f_1f_2$.} \label{tab:U22}
\setlength{\tabcolsep}{0.2cm}
\renewcommand{\arraystretch}{1.4}
\centering
\begin{tabular}{cccccc}
\hline\hline
Phase & $(f_1,f_2)$ & Topological order & $\sigma^{xy}$ & $\kappa^{xy}$ \\ \hline
$U(2)_2$ & $C=2$ & $SU(2)^s_{-2}\cong U(2)_2$ & $1$ & $5/2$ \\ \hline
Laughlin state & $C=1$ & $SU(2)^s_{-1}\cong U(1)_2$ & $1/2$ & $1$ \\ \hline
Mott insulator & $C=0$ & $SU(2)^s_0$ & $0$ & $0$ \\ \hline
Laughlin state & $C=-1$ & $SU(2)^s_{1}\cong U(1)_{-2}$ & $-1/2$ & $-1$ \\ \hline 
$U(2)_{-2}$ & $C=-2$ & $SU(2)^s_{2}\cong U(2)_{-2}$ & $-1$ & $-5/2$ \\ \hline \hline
\end{tabular}
\end{table*}

\begin{table*}[htb]
\caption{Transition out of Ising topological order can be described by the parton construction $S^-=f_1f_2 \phi$.} \label{tab:ITO}
\setlength{\tabcolsep}{0.2cm}
\renewcommand{\arraystretch}{1.4}
\centering
\begin{tabular}{cccccc}
\hline\hline
Phase & $(f_1,f_2)$ & $\phi$ & Topological order & $\sigma^{xy}$ & $\kappa^{xy}$ \\ \hline
Ising topological order & $C=2$ & $\nu=-1/2$ & $U(2)_{2,-2}$ & $-1$ & $1/2$ \\ \hline 
Superfluid & $C=1$ & $\nu=-1/2$ & $ \textrm{Trivial}$ &  &  \\ \hline 
Mott insulator & $C=0$ & $\nu=-1/2$ & $ \textrm{Trivial}$ & $0$ & $0$ \\ \hline 
Laughlin state & $C=-1$ & $\nu=-1/2$ & $U(1)_{-4}$ & $-1/4$ & $-1$ \\ \hline 
Non-Abelian state & $C=-2$ & $\nu=-1/2$ &  $\frac{U(2)_{2,2}^s\times U(1)_{-12}}{Z_2}$ & $-1/3$ & $-5/2$ \\ \hline \hline

\end{tabular}
\end{table*}

\begin{table*}[htb]
\caption{The transition out of anti-Pfaffian state can be described by the parton construction $c=f_1f_2\psi$.} \label{tab:aPf}
\setlength{\tabcolsep}{0.2cm}
\renewcommand{\arraystretch}{1.4}
\centering
\begin{tabular}{cccccc}
\hline\hline
Phase &  $(f_1,f_2)$ & $\psi$   & Topological order & $\sigma^{xy}$ & $\kappa^{xy}$ \\ \hline
anti-Pfaffian & $C=-2$ & $\nu=1/3$ & Anti-Pfaffian & $1/2$ & $-1/2$ \\ \hline
Integer quantum Hall & $C=-1$ & $\nu=1/3$ & Trivial & $1$ & $1$ \\ \hline
Mott insulator & $C=0$ & $\nu=1/3$ & Trivial & $0$ & $0$ \\ \hline 
Laughlin state & $C=1$ & $\nu=1/3$ & $U(1)_5$ & $1/5$ & $1$ \\ \hline 
Non-Abelian state & $C=2$ & $\nu=1/3$ & $\frac{SU(2)_2\times U(1)_{16}}{Z_2}$ & $1/4$ & $5/2$ \\ \hline \hline 
% $f-if$ superconductor & $C=-2$ & $\nu=1$ & confined & & $-3/2$
% \\ \hline \hline 
\end{tabular}
\end{table*}

\begin{table*}[htb]
\caption{The transition out of Wen's $(221)$-parton state can be described by the parton construction $c=f_1f_2\psi$.} \label{tab:221}
\setlength{\tabcolsep}{0.2cm}
\renewcommand{\arraystretch}{1.4}
\centering
\begin{tabular}{cccccc}
\hline\hline
Phase & $(f_1,f_2)$ & $\psi$  & Topological order & $\sigma^{xy}$ & $\kappa^{xy}$ \\ \hline
Wen's $(221)$-parton state & $C=2$ & $\nu=1$ & $U(2)^s_{-2,-4}\cong U(2)_{2,4}$ & $1/2$ & $5/2$ \\ \hline
Laughlin state & $C=1$ & $\nu=1$ & $U(2)_{-1,-3}^s\cong U(1)_3$ & $1/3$ & $1$ \\ \hline
Mott insulator & $C=0$ & $\nu=1$ & $U(2)_{0,-2}^s\cong \textrm{Trivial}$ & $0$ & $0$ \\ \hline
Integer quantum Hall & $C=-1$ & $\nu=1$ & $U(2)_{1,-1}^s\cong U(1)_{-1}$ & $-1$ & $-1$ \\ \hline 
$f-if$ superconductor & $C=-2$ & $\nu=1$ & $U(2)^s_{2,0}$ (confined) & & $-3/2$ \\ \hline  \hline
% $f+if$ superconductor & $C=2$ & $\nu=-1$ & $U(2)^s_{-2,0}$ (confined) & & $3/2$
% \\ \hline \hline
\end{tabular}
\end{table*}

\begin{table*}[htb]
\caption{The phase transition out of Pfaffian state can be described by the parton construction $c=(\psi_1\psi_4-\psi_3\psi_2)\psi/\sqrt{2}$.} \label{tab:Pf}
\setlength{\tabcolsep}{0.2cm}
\renewcommand{\arraystretch}{1.4}
\centering
\begin{tabular}{ccccccc}
\hline\hline
Phase & $(\psi_1,\psi_2)$ & $(\psi_4,\psi_3)$ &$\psi$  & Topological order & $\sigma^{xy}$ & $\kappa^{xy}$ \\ \hline
Pfaffian & $C_1=1$ & $C_2=1$ & $\nu=1$ & Pfaffian & $1/2$ & $3/2$ \\ \hline
Mott insulator & $C_1=0$ & $C_2=0$ & $\nu=1$ & Trivial & $0$ & $0$ \\ \hline
$p-ip$ superconductor & $C_1=-1$ & $C_2=-1$ & $\nu=1$ & Confined & & $-1/2$ \\ \hline 
Mott insulator & $C_1=0$ & $C_2=1$ & $\nu=1$ & Trivial & $0$ & $0$ \\ \hline 
Abelian state & $C_1=-1$ & $C_2=1$ & $\nu=1$ & $K=
\begin{pmatrix} 
0&2&0 \\ 
2&0&1   \\
0&1&1
\end{pmatrix}$ & $1$ & $1$ \\ \hline
Abelian state & $C_1=-2$ & $C_2=1$ & $\nu=1$ & $K=
\begin{pmatrix} 
-4&1 \\ 
1&1   
\end{pmatrix}$ & $4/5$ & $0$ \\ \hline
Non-Abelian state & $C_1=-3$ & $C_2=1$ & $\nu=1$ & $\frac{SU(2)_{-2}\times U(1)_{-16}}{Z_2}\times U(1)_1$   & $3/4$ & $-3/2$ \\ \hline \hline
% $p+ip$ superconductor & $C_1=1$ & $C_2=1$ & $\nu=-1$ & Confined & & $1/2$
% \\ \hline\hline
\end{tabular}
\end{table*}

\begin{table*}[htb]
\caption{The phase transition out of PH-Pfaffian state can be described by the parton construction $c=(\psi_1\psi_4-\psi_3\psi_2)\psi/\sqrt{2}$.} \label{tab:PHPf}
\setlength{\tabcolsep}{0.2cm}
\renewcommand{\arraystretch}{1.4}
\centering
\begin{tabular}{ccccccc}
\hline\hline
Phase &  $(\psi_1,\psi_2)$ & $(\psi_4,\psi_3)$ &$\psi$   & Topological order & $\sigma^{xy}$ & $\kappa^{xy}$ \\ \hline
PH-Pfaffian & $C_1=-1$ & $C_2=-1$ & $\nu=1/3$ & PH-Pfaffian & $1/2$ & $1/2$ \\ \hline 
Mott insulator & $C_1=0$ & $C_2=0$ & $\nu=1/3$ & Trivial & $0$ & $0$ \\ \hline
Non-Abelian state & $C_1=1$ & $C_2=1$ & $\nu=1/3$ & Non-Abelian~\footnote{This Non-Abelian state is described by a Lagrangian
$\mathcal{L}=\frac{1}{4\pi}\mathrm{Tr}(\beta d\beta+\frac{2}{3}\beta^3)+\frac{4}{4\pi}b d b+\frac{1}{2\pi}a d (2b+A)  
    -\frac{3}{4\pi}a d a+4\mathrm{CS}_g$, with an additional cocycle condition between the $USp(4)$ spin gauge field $\beta$ and $U(1)$ gauge field $b$: $\oint(w_2(\beta)/2+db/2\pi)\in \mathbb{Z}$. 
}
& $1/4$ & $3/2$ \\ \hline 
Mott insulator & $C_1=0$ & $C_2=-1$ & $\nu=1/3$ & Trivial & $0$ & $0$ \\ \hline 
Abelian state & $C_1=1$ & $C_2=-1$ & $\nu=1/3$ & $K=
\begin{pmatrix} 
3&1&0 \\ 
1&0&2   \\
0&2&0
\end{pmatrix}
$  & $1/3$ & $1$ \\ \hline
Abelian state & $C_1=2$ & $C_2=-1$ & $\nu=1/3$ & $K=
\begin{pmatrix} 
4&1 \\ 
1&3
\end{pmatrix}
$ & $4/11$ & $2$ \\ \hline 
non-Abelian state & $C_1=3$ & $C_2=-1$ & $\nu=1/3$ & Non-Abelian~\footnote{This Non-Abelian state is described by a Lagrangian
$\mathcal{L}=\frac{2}{4\pi}\mathrm{Tr}(\chi d\chi+\frac{2}{3}\chi^3)+\frac{4}{2\pi}\mathrm{Tr}\chi d b+\frac{4}{4\pi}b d b+\frac{1}{2\pi}a d (2b+A)  
    -\frac{3}{4\pi}a d a+4\mathrm{CS}_g$, with an additional cocycle condition between the $U(2)$ spin gauge field $\chi$ and $U(1)$ gauge field $b$: $\oint(w_2(\chi)/2+(1/2)(d \mathrm{Tr}\chi)/(2\pi)+db/2\pi)\in \mathbb{Z}$. 
} & $3/8$ & $7/2$ \\ \hline  \hline

\end{tabular}
\end{table*}

Similar to the previous example of the bosonic Pfaffian, phase transitions out of other Pfaffian-like states can be described by changing the Chern number of partons.
In Table.\ref{tab:ITO}-\ref{tab:221} we have listed the parton states of various parton constructions for the aforementioned Pfaffian-like states: $U(2)_2$ bosonic state (Table~\ref{tab:U22}), Ising topological order (Table~\ref{tab:ITO}), anti-Pfaffian state (Table~\ref{tab:aPf}), Wen's $(221)$-parton state (Table~\ref{tab:221}), Pfaffian state (Table~\ref{tab:Pf}), and PH-Pfaffian state (Table~\ref{tab:PHPf}).
Among all the parton states, we would like to say a bit more about the topological superconductors.

There are several ways to use partons to construct topological superconductors, which include the $f-if$ superconductor in Table~\ref{tab:221} and the $p-ip$ superconductor in Table~\ref{tab:Pf}. 
The $f-if$ superconductor is constructed via the parton decomposition $c=f_1f_2\psi$, with $(f_1,f_2)$ on a $C=-2$ Chern band and $\psi$ on a $C=1$ Chern band.
The effective theory of this parton construction is a $U(2)^s$ Chern-Simons theory,
\begin{equation}
% \mathcal L=
-\frac{2}{4\pi}\textrm{Tr}(\alpha d \alpha +\frac{2}{3} \alpha^3) + \frac{1}{4\pi}(-\textrm{Tr}\alpha + A) d (-\textrm{Tr}\alpha + A)-3\gcs.
\end{equation}
Here $\alpha$ is a $U(2)$ spin gauge field, $A$ is the probe field (i.e. the electromagnetic field).
To understand why it describes a superconductor, we decompose the $U(2)$ gauge field into a $SU(2)$ and a $U(1)$ gauge field, $\alpha=\alpha_{SU(2)} + \frac{1}{2} (\textrm{Tr}\alpha) \mathbf{1}_2$, so that the effective theory is
\begin{align}
% \mathcal L=
&-\frac{2}{4\pi}\textrm{Tr}[\alpha_{SU(2)} d \alpha_{SU(2)} +\frac{2}{3} \alpha^3_{SU(2)}] -\frac{1}{2\pi} (\textrm{Tr} \alpha) d A \nonumber \\ & +\frac{1}{4\pi} A d A-3\gcs.
\end{align}
Due to the absence of the Chern-Simons term of the $U(1)$ gauge field--$\textrm{Tr}\alpha/2$, the monopole of $\textrm{Tr}\alpha/2$ will condense which confines the $U(1)$ gauge field.
On the other hand, the monopole carries a $q=2$ charge of $A$ (hence it is a Cooper pair), the condensation of it will lead to a superconductor.
Also we note that the gauge structure of the theory is $U(2)=\frac{SU(2)\times U(1)}{Z_2}$, the $SU(2)$ gauge fundamental has to carry a charge of the $U(1)$ gauge field (hence it is a $U(2)$ fundamental).
This means the confinement of the $U(1)$ gauge field will  lead to the confinement of the $SU(2)$ gauge field, even though the latter has a non-trivial Chern-Simons term.
Indeed the $U(2)$ fundamental (i.e. $(f_1,f_2)$) is the vortex of the superconductor, and it has a non-Abelian statistics captured by the $SU(2)^s$ Chern-Simons term.
Moreover, by integrating out $\alpha_{SU(2)}$, we can find that the superconductor has a  thermal Hall conductance $\kappa^{xy}=-3/2$, so it is a $f-if$ superconductor.
With very similar analysis one can also construct a $p-ip$ superconductor with the parton construction shown in Table~\ref{tab:Pf}.

Below we will briefly discuss the phase transitions of all these Pfaffian-like states.

\subsubsection{$U(2)_2$ bosonic state}

The parton decomposition $S^-=f_1f_2$ can be used to construct the $U(2)_2$ bosonic state. 
This parton construction has an emergent $SU(2)$ gauge symmetry, with $(f_1,f_2)^T$ being a $SU(2)$ fundamental.
The $U(2)_2$ state is realized by putting $(f_1,f_2)$ on a $C=2$ Chern band, and by changing the Chern number $C$ of $(f_1, f_2)$ we can obtain various states as summarized in Table~\ref{tab:U22}.
Their transitions are described by,
\begin{itemize}
    \item $N_f=|C-2|$ flavor of Dirac fermions coupled to a $SU(2)_{-(C+2)/2}^s$ Chern-Simons gauge field. 
\end{itemize}

\subsubsection{Ising topological order}

The parton decomposition $S^-=f_1f_2 \phi$ can be used to construct the Ising topological order. 
This parton construction has an emergent $U(2)\cong\frac{SU(2)\times U(1)}{Z_2}$ gauge symmetry, with $(f_1,f_2)^T$ being a $U(2)$ fundamental and $\phi$ being charged under the diagonal $U(1)$ of the $U(2)$ gauge field.
The Ising topological order is realized by putting $(f_1, f_2)$ on a $C=2$ Chern band, and putting $\phi$ into a $\nu=-1/2$ bosonic Laughlin state.
By changing the Chern number $C$ of $(f_1, f_2)$ we can obtain various states as summarized in Table~\ref{tab:ITO}.
The phase transitions between the Ising topological order and other states are described by a $U(2)^s$ gauge theory,
\begin{align}
\mathcal L = &\sum_{I=1}^{N_f=|C-2|} \bar \psi_I (i\slashed \partial + \slashed \alpha) \psi_I+ \frac{1}{4\pi} \frac{C+2}{2}\textrm{Tr}(\alpha d \alpha + \frac{2}{3}\alpha^3) \nonumber \\ 
&+ \frac{2}{4\pi} b d b +\frac{1}{2\pi} b d (A-\textrm{Tr} (\alpha)).
\end{align}
Here $\alpha$ is a $U(2)$ spin gauge field, $b$ is $U(1)$ gauge field which describes the $\nu=-1/2$ Laughlin state of $\phi$, $A$ is a $U(1)$ background field. 
We omit the gravitational Chern-Simons term as well as the mass of Dirac fermions for convenience.

We also remark that some phase transitions can be described by changing the state formed by $\phi$.
For example, if $\phi$ forms a superfluid state, the corresponding parton state is the $U(2)_2$ bosonic state, hence its transition is a gauged version of the superfluid--$\nu=-1/2$ Laughlin state transition~\cite{Maissam2014_FQHtransition,PhysRevX.8}\footnote{In the composite fermion approach, this transition can be described as the transition of topological superconductors formed by composite fermion, whose critical theory is four Majorana fermions coupled to a $Z_2$ gauge field. With few lines of algebra, one can show that the two different approaches are indeed equivalent.}.

\subsubsection{Anti-Pfaffian state}

The parton decomposition $c=f_1f_2 \psi$ can be used to construct the anti-Pfaffian state. Similar to the parton construction for the Ising topological order, this construction has an emergent $U(2)$ gauge symmetry, with $(f_1,f_2)^T$ being a $U(2)$ fundamental and $\psi$ being charged under the diagonal $U(1)$ of the $U(2)$ gauge field.
The anti-Pfaffian state is realized by putting $(f_1,f_2)$ on a $C=-2$ Chern band, meanwhile $\psi$ is forming a $\nu=1/3$ Laughlin state. 
By changing the Chern number $C$ of $(f_1, f_2)$ we can obtain various states as summarized in Table~\ref{tab:aPf}.
The phase transitions between the anti-Pfaffian state and other partons states are described by a $U(2)^s$ gauge theory,
\begin{align}
\mathcal L = &\sum_{I=1}^{N_f=|C+2|} \bar \psi_I (i\slashed \partial + \slashed \alpha) \psi_I+ \frac{1}{4\pi} \frac{C-2}{2}\textrm{Tr}(\alpha d \alpha + \frac{2}{3}\alpha^3) \nonumber \\ 
&-\frac{3}{4\pi} b d b +\frac{1}{2\pi} b d (A-\textrm{Tr} (\alpha)).
\end{align}
Here $\alpha$ is a $U(2)$ spin gauge field, $b$ is $U(1)$ gauge field which describes the $\nu=1/3$ Laughlin state of $\psi$, $A$ is a $U(1)$ probe field (i.e. the electromagnetic field).

\subsubsection{Wen's $(221)$-parton state}

The parton decomposition $c=f_1f_2 \psi$ can also be used to construct Wen's $(221)$-parton state.
The difference from the anti-Pfaffian state is that, here we put $(f_1,f_2)$ on a $C=2$ Chern band, and $\psi$ on a $C=1$ Chern band.
By changing the Chern number $C$ of $(f_1, f_2)$ we can obtain various states as summarized in Table~\ref{tab:221}.
The phase transitions between the Wen's $(221)$-parton state and other partons states are described by a $U(2)^s$ gauge theory,
\begin{align}
\mathcal L = &\sum_{I=1}^{N_f=|C-2|} \bar \psi_I (i\slashed \partial + \slashed \alpha) \psi_I+ \frac{1}{4\pi} \frac{C+2}{2}\textrm{Tr}(\alpha d \alpha + \frac{2}{3}\alpha^3) \nonumber \\ 
&-\frac{1}{4\pi} b d b +\frac{1}{2\pi} b d (A-\textrm{Tr} (\alpha)).
\end{align}
Here $\alpha$ is a $U(2)$ spin gauge field, $b$ is $U(1)$ gauge field which describes the $\nu=1$ integer quantum Hall state of $\psi$, $A$ is a $U(1)$ probe field (i.e. the electromagnetic field).

Also one can change the Chern number of $\psi$ instead of $(f_1,f_2)$.
For example, if $\psi$ realizes a $C=-1$ Chern insulator, the parton state will be a $f+if$ superconductor.
The phase transition will then be described by a QED$_3$-Chern-Simons theory rather than a QCD$_3$-Chern-Simons theory.

\subsubsection{Pfaffian state}

The parton decomposition $c=(\psi_1\psi_4-\psi_3\psi_2)\psi/\sqrt{2}$ can be used to construct the Pfaffian state.
The maximal gauge symmetry of this parton construction is $\frac{USp(4)\times U(1)}{Z_2}$, and $(\psi_1,\psi_2,\psi_3,\psi_4)$ is the bi-fundamental of  $USp(4)$ and $U(1)$, while $\psi$ carries $q=2$ charge of $U(1)$ but neutral under $USp(4)$.
The Pfaffian state is realized by putting $(\psi_1,\psi_2,\psi_3,\psi_4)$ in a $C=1$ Chern band, and $\psi$ in a $C=1$ Chern band as well.
Due to the $USp(4)^s$-$U(2)^s$ duality, we can also higgs the $USp(4)$ down to $U(2)$ without destroying the Pfaffian state.
In this case, the emergent gauge symmetry is  $\frac{U(2)\times U(1)}{Z_2}$. $(\psi_1,\psi_2)$ and $(\psi_4, \psi_3)$ are the $U(2)$ fundamental and anti-fundamental, respectively.
Also $(\psi_1,\psi_2)$ and $(\psi_4, \psi_3)$ carry unit charge of the $U(1)$ gauge field, and $\psi$ carries $q=2$ gauge charge of the $U(1)$ gauge field.

Table~\ref{tab:Pf} summarizes several parton states for different choices of Chern number $C_1$ (of $(\psi_1,\psi_2)$) and $C_2$ (of $(\psi_4, \psi_3)$).
If we keep the maximal $\frac{USp(4)\times U(1)}{Z_2}$ gauge symmetry, we shall have $C_1=C_2=C$.
This structure can be used to describe phase transitions from the Pfaffian to the Mott insulator or $p-ip$ superconductor.
The critical theory is,
\begin{align}
&\mathcal L = \sum_{I=1}^{N_f=|C-1|} \bar \psi_I (i\slashed \partial + \slashed \alpha+\slashed b \mathbf{1}_4) \psi_I + \frac{2(C+1)}{4\pi} b d b \nonumber \\ 
& + \frac{1}{4\pi} \frac{C+1}{2}\textrm{Tr}(\alpha d \alpha + \frac{2}{3}\alpha^3) - \frac{1}{4\pi} c d c +\frac{1}{2\pi} c d (A-2b).
\end{align}
Here $\alpha$ is a $USp(4)$ spin gauge field, $b$ is a $U(1)$ gauge field.
$c$ is a $U(1)$ gauge field which describes the  Chern number $1$ insulator of $\psi$, $A$ is a $U(1)$ probe field (i.e. the electromagnetic field). Note that the $\psi_i$ $(i=1,2,3,4)$ partons couple to an ordinary $U(1)$ gauge field $b$, since the "fermionic core" of Wilson lines in half-integer spin representations of $USp(4)$ has already been taken into account by $\alpha$.  
We omit the gravitational Chern-Simons term as well as the mass of Dirac fermions for convenience.

We can also break the $\frac{USp(4)\times U(1)}{Z_2}$ gauge symmetry down to $\frac{U(2)\times U(1)}{Z_2}$.
For this gauge structure, $C_1$ and $C_2$ has to be tuned independently (we consider the case of tuning $C_1$). 
It can be used to describe different phase transitions from the ones discussed above.
The critical theory is, 
\begin{align}
\mathcal L = &\sum_{I=1}^{N_f=|C_1-1|} \bar \psi_I (i\slashed \partial + \slashed \chi+\slashed b \mathbf{1}_2) \psi_I+\frac{C_1+3}{4\pi}b db \nonumber \\ &+  \frac{1}{4\pi} \frac{C_1+3}{2}\textrm{Tr}(\chi d \chi + \frac{2}{3}\chi^3) + \frac{C_1-1}{4\pi} (\textrm{Tr} \chi) db \nonumber \\ 
&- \frac{1}{4\pi} c d c +\frac{1}{2\pi} c d (A-2b).
\end{align}
Here $\chi$ is a $U(2)$ spin gauge field, $b$ is a $U(1)$ gauge field.
$c$ is a $U(1)$ gauge field which describes the $\nu=1$ Chern insulator of $\psi$, $A$ is a $U(1)$ probe field (i.e. the electromagnetic field). 
We omit the gravitational Chern-Simons term as well as the mass of Dirac fermions for convenience.

At last we remark that some phase transitions can be described by changing the state formed by the parton $\psi$.
For example, if $\psi$ forms a $\nu=-1$ integer quantum Hall state the parton construction will then yield the $p+ip$ superconductor. 
The transition between the $p+ip$ superconductor and the Pfaffian is then described by a QED$_3$-Chern-Simons theory, namely $N_f=2$ Dirac fermions coupled to a $U(1)$ Chern-Simons field (i.e. $2b$).
One can also use the composite fermion approach to describe this phase transition, and it is indeed equivalent to the QED$_3$-Chern-Simons theory described above. 
In constrast, the transition between the $p-ip$ superconductor and Pfaffian state can only be described by the QCD$_3$-Chern-Simons theory constructed by the $USp(4)$ parton construction. 

\subsubsection{PH-Pfaffian state}

Similar to the Pfaffian state, the PH-Pfaffian state can also be constructed via the parton decomposition $c=(\psi_1\psi_4-\psi_3\psi_2)\psi/\sqrt{2}$.
The PH-Pfaffian state is realized by putting $(\psi_1,\psi_2,\psi_3,\psi_4)$ in a $C=-1$ Chern band, and $\psi$ in a $\nu=1/3$ Laughlin state.
The PH-Pfaffian state is well captured by either the maximal $\frac{USp(4)\times U(1)}{Z_2}$ or a smaller $\frac{U(2)\times U(1)}{Z_2}$ gauge symmetry.
Table~\ref{tab:PHPf} summarizes various parton states of different Chern number assignment to the fermionic partons.
The phase transitions between the PH-Pfaffian state and other states are described by either the $\frac{USp(4)\times U(1)}{Z_2}$ or $\frac{U(2)\times U(1)}{Z_2}$ gauge depending on whether $C_1$ and $C_2$ are changing simultaneously or independently.
The $\frac{USp(4)\times U(1)}{Z_2}$ gauge theory is,
\begin{align}
&\mathcal L = \sum_{I=1}^{N_f=|C+1|} \bar \psi_I (i\slashed \partial + \slashed \alpha+\slashed b \mathbf{1}_4) \psi_I + \frac{2(C-1)}{4\pi} b d b \nonumber \\ 
& + \frac{1}{4\pi} \frac{C-1}{2}\textrm{Tr}(\alpha d \alpha + \frac{2}{3}\alpha^3) - \frac{3}{4\pi} c d c +\frac{1}{2\pi} c d (A-2b).
\end{align}
Here $\alpha$ is a $USp(4)$ spin gauge field, $b$ is a $U(1)$ gauge field.
$c$ is a $U(1)$ gauge field which describes the $\nu=1/3$ Laughlin state of $\psi$, $A$ is a $U(1)$ probe field (i.e. the electromagnetic field). 

The $\frac{U(2)\times U(1)}{Z_2}$ critical theory is, 
\begin{align}
\mathcal L = &\sum_{I=1}^{N_f=|C_1+1|} \bar \psi_I (i\slashed \partial + \slashed \chi+\slashed b \mathbf{1}_2) \psi_I+\frac{C_1-3}{4\pi}b d b \nonumber \\ &+  \frac{1}{4\pi} \frac{C_1-3}{2}\textrm{Tr}(\chi d \chi + \frac{2}{3}\chi^3) + \frac{C_1+1}{4\pi} (\textrm{Tr} \chi) d b \nonumber \\ 
&- \frac{3}{4\pi} c d c +\frac{1}{2\pi} c d (A-2b).
\end{align}
Here $\chi$ is a $U(2)$ spin gauge field, $b$ is a $U(1)$ gauge field.
$c$ is a $U(1)$ gauge field which describes the $\nu=1$ Chern insulator of $\psi$, $A$ is a $U(1)$ probe field (i.e. the electromagnetic field).

\section{\label{sec:sum} Summary and discussion}

We study a new universality class of phase transitions of FCIs/FQHs.
These transitions have emergent non-Abelian gauge fields coupled to $N_f$ flavors of Dirac fermions, hence are dubbed QCD$_3$-Chern-Simons theory.
In contrast to the previous study of the  QED$_3$-Chern-Simons theory~\cite{PhysRevX.8}, the transitions we consider here are either 1) between Abelian FCIs/FQHs in different Jain sequences, or 2) involving non-Abelian FCIs/FQHs.  
Specifically, we use the non-Abelian parton construction to construct these FCIs/FQHs, and the transitions between them are nothing but the Chern number changing transitions of non-Abelian partons.
In order to corretly characterize the topological order from parton constructions, we utilize the level-rank duality~\cite{hsin2016level}, and clarify its meaning for the condensed matter application.
For the transitions involving non-Abelian states, we focus on the Pfaffian-like states, but the generalization to other more complicated non-Abelian states is straightforward.
We shall also mention that the phase transitions discussed here and previously~\cite{PhysRevX.8} are far from a complete list.
For example, a recent numerically observed phase transition~\cite{Schoonderwoerd2019} falls beyond our study.

We remark that in this paper we only discuss effective theories of these phase transitions without delving into their dynamics (i.e. RG flow, critical exponents, etc.).
It is possible that when the Dirac flavor $N_f$ is small, the effective theory may not flow into a nontrivial conformal fixed point (in other words it will be weakly first order phase transitions).
If this is the case for certain theories, interesting critical phenomenon can still appear at finite temperature due to the proposed scenario of complex fixed point (also dubbed pseudo-criticality )~\cite{wang2017deconfined,gorbenko2018walking,gorbenko2018walking2}.
We hope this new proposed scenario will also motivate more experimental study on these phase transitions.

\begin{acknowledgments}
We thank Chong Wang and Liujun Zou for stimulating discussions. 
Research at Perimeter Institute is supported in part by the Government of Canada through the Department of Innovation, Science and Economic Development Canada and by the Province of Ontario through the Ministry of Colleges and Universities.
R. M. acknowledges support from NSERC Discovery Grant 50503-11018. 
\end{acknowledgments}

\appendix

\section{\label{app:duality}Duality between different Chern-Simons theories}

\subsection{Pfaffian state}
The fermionic Pfaffian state is described by the following TQFT~\cite{lian2018theory}
\begin{equation}
\frac{\mathrm{Ising}\times U(1)_8}{Z_2}\sim \frac{U(2)_{2,-2}\times U(1)_8}{Z_2}.
\end{equation}
The Lagrangian is
\begin{align}
\mathcal L=&
-\frac{2}{4\pi}\mathrm{Tr}(\tilde{b} d \tilde{b}+\frac{2}{3}\tilde{b}^3)+\frac{2}{4\pi}(\mathrm{Tr}\tilde{b}) d (\mathrm{Tr}\tilde{b}) \nonumber \\
&-\frac{8}{4\pi}\tilde{c} d\tilde{c}+\frac{2}{2\pi}Ad\tilde{c},
\label{eq:Pfabeforequotient}
\end{align}
where $\tilde{b}$ is a $U(2)$ gauge field, and $\tilde{c}$ is a $U(1)$ gauge field.
$\tilde{b}$ and $\tilde{c}$ are not independent gauge field as one needs to mod out their $Z_2$ center.
$A$ is the background electromagnetic field, and we need a factor 2 in the last term to reproduce the correct filling fraction $\nu=\frac{1}{2}$ for the Pfaffian state. 
From high energy point of view, since this theory Eq.~(\ref{eq:Pfabeforequotient}) (without modding out the $Z_2$ center) is a non-spin TQFT  due to the Chern-Simons levels, we need this factor of 2 if we interpret $A$ as a background $\mathrm{spin}_c$ connection. 

The procedure of modding out $Z_2$ center of $\tilde{b}$ and $\tilde{c}$ can be done by gauging a $Z_2$ one-form global symmetry. As discussed in Ref.~\cite{seiberg2016gapped}, one can achieve this by a change of variables to conventionally normalized gauge fields, such that the $Z_2$ one-form symmetry does not act on them. First we note that if $\tilde{b}$ and $\tilde{c}$ are independent fields, we should have 
\begin{equation}
\oint \left(\frac{1}{2}\frac{d\mathrm{Tr} \tilde{b}}{2\pi}+\frac{w_2(\tilde{b})}{2}\right)\in \mathbb{Z}, \quad \oint \left(\frac{d\tilde{c}}{2\pi}\right)\in \mathbb{Z}.
\end{equation}
The above two equations means nothing but that only $2\pi$ flux of $\tilde{b}$ or $\tilde{c}$ is allowed.
Next moding out the $Z_2$ center of $\tilde{b}$ and $\tilde{c}$ means that: we allow $\pi$ flux of $\tilde{b}$ and $\tilde{c}$ on a closed surface, but we need to identify them,
\begin{align}
&\oint \left(\frac{1}{2}\frac{d\mathrm{Tr}\tilde{b}}{2\pi}+\frac{w_2(\tilde{b})}{2}\right)\in \frac{1}{2} \mathbb{Z}, \, \, \oint \left(\frac{d\tilde{c}}{2\pi}\right)\in \frac{1}{2} \mathbb{Z}, \\
&\oint \left( \frac{1}{2}\frac{d\mathrm{Tr}\tilde{b}}{2\pi}+\frac{w_2(\tilde{b})}{2}-\frac{d\tilde{c}}{2\pi}\right)\in \mathbb{Z}.
\end{align}

To solve above constraints, we can redefine gauge fields,
\begin{align}
b &=\tilde b-\tilde c\textbf{1}_2,\\
 c&=2\tilde{c}.
\end{align}
One can find that the constraints reduced to that only $2\pi$ gauge flux of $b$  and $c$ is allowed.

In terms of these new variables the Lagrangian can be written as 
\begin{align}
\mathcal L =&
-\frac{2}{4\pi}\mathrm{Tr}(b d b+\frac{2}{3}b^3)+\frac{2}{4\pi}(\mathrm{Tr}b) d(\mathrm{Tr}b)\nonumber\\
&+\frac{1}{2\pi}c d(A-\Tr b)-\frac{1}{4\pi}cdc.
\label{eq:pfbosonic}
\end{align}
This is the Chern-Simons theory description for Pfaffian state.
Indeed we can integrate out the $U(1)$ gauge field $c$, yielding
\begin{equation}
\begin{split}
\mathcal L = & -\frac{2}{4\pi}\mathrm{Tr}(b d b+\frac{2}{3} b^3)+\frac{3}{4\pi}(\mathrm{Tr}b) d(\mathrm{Tr}b) \\ 
& - \frac{1}{2\pi} A d \Tr b + \frac{1}{4\pi} A d A + \gcs.
\end{split}
\end{equation}
This is nothing but the particle-hole conjugate of anti-Pfaffian ($U(2)_{-2,4}$).

Now we try to find a dual spin gauge field description for the Pfaffian state. From our previous result Eq.(\ref{eq:Isingspin}) and $USp(4)^s-U(2)^2$ level-rank duality, the Pfaffian topological order can be rewritten as 
\begin{equation}
       \mathrm{Pfaffian}\cong\frac{\frac{USp(4)_{-1}^s\times U(1)_{-4}}{Z_2}\times U(1)_8}{Z_2}.
       \label{eq:pffermion}
\end{equation}
To construct a Lagrangian, we start from two decoupled theories, namely $\frac{USp(4)_{-1}^s\times U(1)_{-4}}{Z_2}$ and $U(1)_8$, and then implement the $Z_2$ quotient. The Lagrangian for $\frac{USp(4)_{-1}^s\times U(1)_{-4}}{Z_2} \times U(1)_8$ theory is 
\begin{equation}
    \frac{1}{4\pi}\mathrm{Tr}(\beta d\beta+\frac{2}{3}\beta^3)+\frac{4}{4\pi}\tilde{a} d\tilde{a} 
    -\frac{8}{4\pi}\tilde{b} d \tilde{b}+\frac{2}{2\pi}Ad\tilde{b},
 \end{equation}
 with a cocycle condition
 \begin{equation}
     \oint(\frac{w_2(\beta)}{2}+\frac{d\tilde{a}}{2\pi})\in \mathbb{Z}.
 \end{equation}
 Here $w_2(\beta)\in H^2(M,Z_2)$ is the second Stiefel-Whitney class of the $USp(4)^s$ bundle~\cite{nakahara2003geometry}. The overall $Z_2$ quotient on the denominator Eq.(\ref{eq:pffermion}) imposes a constraint on the normalization of the gauge fields
 \begin{equation}
     \oint \left( \frac{w_2(\beta)}{2}+\frac{d\tilde{a}}{2\pi}-\frac{d\tilde{b}}{2\pi}\right)\in \mathbb{Z}.
 \end{equation}
  The above constraint should be solved by a change of variables to conventionally normalized $\frac{USp(4)^s\times U(1)}{Z_2}$ and $U(1)$ gauge fields, such that the $Z_2$ one-form global symmetry we try to gauge does not act on the new variables:
 \begin{equation}
     \begin{split}
      a=&2(\tilde{a}-\tilde{b}),\\
      b=&2\tilde{b}.
     \end{split}
 \end{equation}
This change of variable is chosen to satisfy the two following cocycle conditions:
\begin{align}
       &\oint(\frac{w_2(\beta)}{2}+\frac{1}{2}\frac{da}{2\pi})\in \mathbb{Z},
       \label{eq:cocyclePfa}
       \\
       &\oint \frac{db}{2\pi}\in \mathbb{Z}.
\end{align}
 These two conditions ensure that $b$ is a good $U(1)$ gauge field and $(\beta+\frac{a}{2}\mathbf{1}_4)$ is a good $\frac{USp(4)^s\times U(1)}{Z_2}$ gauge field. In terms of these new variables the Lagrangian becomes
 \begin{equation}
      \frac{1}{4\pi}\mathrm{Tr}(\beta d\beta+\frac{2}{3}\beta^3)+\frac{1}{4\pi}a da 
    -\frac{1}{4\pi}b d b+\frac{1}{2\pi}(A+a) db.
 \end{equation}
Including the gravitational term arising, we can get the spin gauge field description of the Pfaffian state in Eq. (\ref{eq:pffermionusp}).
To see the topological order, ne can set $b=c+a$ in which $c$ is a conventionally normalized $U(1)$ gauge field,
\begin{equation}
\mathcal L_{pf}=\frac{1}{4\pi}\mathrm{Tr}(\beta d\beta+\frac{2}{3}\beta^3)+\frac{2}{4\pi}a d a-\frac{1}{4\pi}c dc+\frac{1}{2\pi}(a+c) dA.
\end{equation}
Together with the constraint Eq.~(\ref{eq:cocyclePfa}), one can see the gauge group is $\frac{USp(4)_{-1}^s\times U(1)_{-8}}{Z_2}\times U(1)_1$~\footnote{Here the $U(1)_{-8}$ sector is $\frac{8}{4\pi}(\frac{a}{2})d(\frac{a}{2})$, since the elementary gauge charge of $a/2$ is quantized as $1$. The TQFT can be consistently put on $\mathrm{spin}_c$ manifold. However, the Lagrangian is written in terms of $a$ since it has conventional flux quantization: the flux on any closed surfaces is $2\pi$ integer.} and the quasi-particles in fundamental representation of the $USp(4)$ gauge group should carry charge $\frac{1}{2}$ under $a$ (or charge 1 under $a/2$), see Eq.~(\ref{eq:Integratepartonpf}). Our result is the same as that in Ref.~\cite{lian2018theory}, which states that the Pfaffian topological order can be described by $(SU(2)_2\times U(1)_{-8})/Z_2\times U(1)_1$ Chern-Simons theory.

\subsection{Anti-Pfaffian state}

The topological order of Anti-Pfaffian state is known to be described by a Chern-Simons gauge theory \cite{lian2018theory}
\begin{equation}
    \frac{SU(2)_{-2}\times U(1)_8}{Z_2}.
\end{equation}
Using the basic level/rank duality, it is straightforward to see that a spin gauge field description for Anti-Pfaffian topological order is
\begin{equation}
    \frac{U(2)_{2,2}^s\times U(1)_8}{Z_2}.
\end{equation}
In order to get a concrete Lagrangian, we start from two decoupled theories, namely $U(2)_{2,2}^s$ and $U(1)_8$, and then gauge a common $Z_2$ one form symmetry \cite{gaiotto2015generalized}. 

The Lagrangian for $U(2)_{2,2}^s\times U(1)_8$ theory is 
\begin{equation}
    -\frac{2}{4\pi}\mathrm{Tr}(\tilde{\chi} d\tilde{\chi}+\frac{2}{3}\tilde{\chi}^3)-\frac{8}{4\pi}\tilde{c} d \tilde{c}+\frac{2}{2\pi}\tilde{c} dA.
\label{eq:AntiPfdecouple}
\end{equation}
The $Z_2$ quotient is now done by finding an appropriate change of variables to conventionally normalized $U(2)^s$ and $U(1)$ gauge fields, such that the $Z_2$ one form global symmetry does not act on the new variables:
\begin{equation}
\begin{split}
\chi&=\tilde{\chi}-\tilde{c}\mathbf{1}_2,\\
c&=2\tilde{c}.
\end{split}
\end{equation}
Physically, in both of the two decoupled theories there is a line operator which generates a $Z_2$ one form symmetry. One can combine the two lines, and the $Z_2$ quotient procedure projects out all the lines which have non-trivial braiding with this combined line. This line, which has spin-$1/2$, is identified with physical electron. Including the gravitational response term arising from the level-rank duality, the Lagrangian of Anti-Pfaffian state in terms of these new variables is 
\begin{equation}
    \mathcal L_{apf}=-\frac{2}{4\pi}\mathrm{Tr}(\chi d\chi+\frac{2}{3}\chi^3)-\frac{3}{4\pi}cdc+\frac{1}{2\pi}cd(\mathrm{Tr}\chi+A)-4\mathrm{CS}_g,
\end{equation}
which is the Chern-Simons gauge theory in Eq.(\ref{eq:apffermionic}).

\subsection{PH-Pfaffian state}
The topological order of PH-Pfaffian state is described by the following Chern-Simons gauge theory \cite{lian2018theory}
\begin{equation}
    \frac{\overline{\mathrm{Ising}}\times U(1)_8}{Z_2}.
\end{equation}
From our previous result Eq.(\ref{eq:Isingspin}) and $USp(4)^s-U(2)^2$ level-rank duality, this topological order can be rewritten as 
\begin{equation}
       \frac{\frac{USp(4)_1^s\times U(1)_{4}}{Z_2}\times U(1)_8}{Z_2}.
       \label{eq:PHpffermion}
\end{equation}
To construct a Lagrangian, we start from three decoupled theories, namely $USp(4)_1^s$, $U(1)_4$ and $U(1)_8$, and then gauge the two corresponding $Z_2$ one form symmetries. The Lagrangian for $USp(4)_1^s\times U(1)_4 \times U(1)_8$ theory is 
\begin{equation}
    -\frac{1}{4\pi}\mathrm{Tr}(\beta d\beta+\frac{2}{3}\beta^3)-\frac{4}{4\pi}\tilde{b} d\tilde{b} 
    -\frac{8}{4\pi}\tilde{a} d \tilde{a}+\frac{2}{2\pi}Ad\tilde{a}.
 \end{equation}
 The overall $Z_2$ quotient in Eq.(\ref{eq:PHpffermion}) imposes a constraint on the normalization of gauge fields
 \begin{equation}
     \oint \left( \frac{w_2(\beta)}{2}+\frac{d\tilde{b}}{2\pi}-\frac{d\tilde{a}}{2\pi}\right)\in \mathbb{Z},
 \end{equation}
 where $w_2(\beta)\in H^2(M,Z_2)$ is the second Stiefel-Whitney class of $\beta$, which could acquire a non-trivial contribution from a twisted $USp(4)$ gauge bundle. 
 The above constraint, which effectively changes the gauge group to $\frac{USp(4)^s\times U(1)\times U(1)}{Z_2}$, should be solved by a change of variables to conventionally normalized gauge fields, such that this overall $Z_2$ quotient does not act on the new variables:
 \begin{equation}
     \begin{split}
      b=&2(\tilde{b}-\tilde{a}),\\
      a=&2\tilde{a}.
     \end{split}
 \end{equation}
This change of variable is chosen to satisfy the two following cocycle conditions imposed by the other $Z_2$ quotient (on the numerator of Eq.(\ref{eq:PHpffermion})):
\begin{align}
       &\oint(\frac{w_2(\beta)}{2}+\frac{1}{2}\frac{db}{2\pi})\in \mathbb{Z},
       \label{eq:cocyclePH}
       \\
       &\oint \frac{da}{2\pi}\in \mathbb{Z}.
\end{align}
 These two conditions ensure that $a$ is a good $U(1)$ gauge field and $(\beta+\frac{b}{2}\mathbf{1}_4)$ is a good $\frac{USp(4)^s\times U(1)}{Z_2}$ gauge field.
 In terms of the new variables, the Lagrangian of PH-Pfaffian state is, after including the gravitational response,  
\begin{align}
\mathcal L_{phpf}=&-\frac{1}{4\pi}\mathrm{Tr}(\beta d\beta+\frac{2}{3}\beta^3)-\frac{1}{4\pi}b db-\frac{1}{2\pi}b da \nonumber\\
&-\frac{3}{4\pi}a d a+\frac{1}{2\pi}Ada-4\mathrm{CS}_g.
\end{align}
Hence we recover the TQFT description of PH-Pfaffian state shown in Eq.(\ref{eq:phfermionic}). Note that the cocycle condition Eq.~(\ref{eq:cocyclePH}) reminds us that the quasi-particles in the fundamental representation of $USp(4)$ should carry charge $\frac{1}{2}$ under $b$, see discussions around Eq.~(\ref{eq:PHparton}).

\bibliography{non_Abelian_FCI}

\end{document}